\documentclass[12pt]{article} 
\pdfoutput=1

\usepackage{amsmath,amssymb,amsfonts}
\usepackage{color}
\definecolor{darkblue}{rgb}{0.1,0.1,.7}
\usepackage[colorlinks, linkcolor=darkblue, citecolor=darkblue, urlcolor=darkblue, linktocpage]{hyperref} 
\usepackage[square, comma, compress,numbers]{natbib}
\usepackage[]{amsmath}
\usepackage[]{graphicx}
\usepackage[]{latexsym}
\usepackage{geometry}
\usepackage{amscd}
\usepackage[all,cmtip]{xy}
\usepackage{mathrsfs}
\usepackage{bbold}
\usepackage{subfigure}
\usepackage[margin=10pt,font=small,labelfont=bf]{caption}
\usepackage{simplewick}
\usepackage{changepage}
\usepackage[]{algorithm2e}
\usepackage{booktabs,multirow}
\definecolor{darkblue}{rgb}{0.,0.,0.4}
\definecolor{darkred}{rgb}{0.5,0.,0.}
\definecolor{BlueViolet}{RGB}{138,43,226}
\definecolor{SkyBlue}{RGB}{30,144,255}
\definecolor{DarkGreen}{RGB}{0,100,0}

\makeatletter
\newsavebox{\@brx}
\newcommand{\llangle}[1][]{\savebox{\@brx}{\(\m@th{#1\langle}\)}%
  \mathopen{\copy\@brx\kern-0.5\wd\@brx\usebox{\@brx}}}
\newcommand{\rrangle}[1][]{\savebox{\@brx}{\(\m@th{#1\rangle}\)}%
  \mathclose{\copy\@brx\kern-0.5\wd\@brx\usebox{\@brx}}}
\makeatother

\newcommand\s{\sigma}
\newcommand\e{\epsilon}
\newcommand\<{\langle}
\renewcommand\>{\rangle}

\newcommand\ds{\Delta_{\s}}
\newcommand\de{\Delta_{\e}}

\begin{document}

\vspace*{-.6in} \thispagestyle{empty}
\vspace{.2in} {\Large
\begin{center}
{\bf 
Precision Bootstrap for the $\mathcal{N}=1$ Super-Ising Model
}
\end{center}
}
\vspace{.2in}
\begin{center}
{\bf 
Alexander Atanasov$^{a}$,
Aaron Hillman$^{b}$,
David Poland$^{c}$,\\
Junchen Rong$^{d}$,
Ning Su$^{e}$
\\
\vspace{.2in} 
$^a$ {\it Center for the Fundamental Laws of Nature, Harvard University, \\Cambridge, MA 02138, USA}\\
$^b$ {\it Department of Physics, Jadwin Hall, Princeton University, NJ, USA 08540}\\
$^c$ {\it Department of Physics, Yale University, New Haven, CT 06520, USA}\\
$^d$ {\it Institut des Hautes \'Etudes Scientifiques, 91440 Bures-sur-Yvette, France }\\
$^e$ {\it Department of Physics, University of Pisa, I-56127 Pisa, Italy}
}
\end{center}

\vspace{.2in}

\begin{abstract}
In this note we report an improved determination of the scaling dimensions and OPE coefficients of the minimal supersymmetric extension of the 3d Ising model using the conformal bootstrap. We also show how this data can be used as input to the Lorentzian inversion formula, finding good agreement between analytic calculations and numerical extremal spectra once mixing effects are resolved.
\end{abstract}

\newpage

\tableofcontents
\newpage

\section{Introduction}
\label{sec:introduction}
The modern conformal bootstrap has yielded the world's most precise critical exponents in the 3d Ising model \cite{Kos:2016ysd}. Another theory that appears to be solvable by bootstrap methods is the 3d $\mathcal{N}=1$ supersymmetric extension of the Ising model, or the $\mathcal{N}=1$ super-Ising model \cite{Rong:2018okz,Atanasov:2018kqw}. The critical exponents of this theory can also be determined to high precision using bootstrap methods. In this paper, we continue the study of this model both numerically and analytically. 

This model is also of relevance to condensed matter physics. In particular, it was argued in \cite{Grover:2013rc} that the corresponding superconformal fixed point can be realized as a quantum critical point at the boundary of a topological superconductor. The $\mathcal{N}=1$ super-Ising model contains only one relevant operator that is invariant under time-reversal symmetry. Physically, such a property means that non-supersymmetric renormalization groups flows can reach the fixed point by just tuning the coupling of the corresponding operator to zero. This property is called ``emergent supersymmetry'' and is critical for experimental realization. 

In the first part of the paper, we push the numerical calculation of the critical exponents of the $\mathcal{N}=1$ Ising model to higher precision. Determining the critical exponents of the $\mathcal{N}=1$ super-Ising model is desirable for many reasons.
This model is the first entry of a family of models called Gross-Neveu-Yukawa models, whose Lagrangian is 
\begin{align}
\mathcal{L}=\frac{1}{2}(\partial_{\mu}\sigma)^2+\sum_{i=1}^{N}\bar{\psi}^{i}\gamma^{\mu}\partial_{\mu}\psi^{i}+ \sum_{i=1}^{N} \frac{1}{2}\lambda_1\sigma  \bar{\psi}^i\psi^i +\frac{1}{8}\lambda_2 \sigma^4.
\end{align}
These models and their variations have many interesting applications in condensed matter physics. 
In particular, a model with $N=8$ fermions describes the quantum phase transition from the semi-metal phase to the charge density instability phase in graphene \cite{Herbut:2006cs}. 
There is a large literature on theoretical studies of these models using perturbative methods (see for example~\cite{Hasenfratz:1991it,Zinn-Justin:1991ksq,Gracey:1990wi,Gracey:1992cp,Gracey:1993kb,Diab:2016spb}), quantum Monte Carlo simulations \cite{Karkkainen:1993ef,Chandrasekharan:2013aya,Li:2014aoa,Hesselmann:2016tvh,Fei:2016sgs,Huffman:2017swn}, and the (non-supersymmetric) numerical bootstrap \cite{Iliesiu:2015qra,Iliesiu:2017nrv,fermionbootstrap2021}. Reproducing the precise value of the critical exponents from the superconformal numerical bootstrap will be an important consistency check on other methods. Knowing the precise critical exponents of the $N$=1 model also allows one to perform a two-sided Pad\'e approximation of the large-$N$ perturbative series \cite{Gracey:1990wi,Gracey:1992cp,Gracey:1993kb} and improves the predictions of critical exponents for models with higher $N$. See~e.g.~\cite{Rong:2018okz}. 

The development of the numerical bootstrap program has been complemented by recent progress in the analytic bootstrap~\cite{Alday:2007mf,Komargodski:2012ek,Fitzpatrick:2012yx,Alday:2016njk,Simmons-Duffin:2016wlq,Caron-Huot:2017vep}.
In particular, the Lorentzian inversion formula~\cite{Caron-Huot:2017vep, Simmons-Duffin:2017nub} allows one to make precise predictions for the conformal data of operators that belong to the low twist  Regge trajectories, using the scaling dimensions and OPE coefficients of low-dimension operators as input~\cite{Albayrak:2019gnz, Liu:2020tpf, Caron-Huot:2020ouj}. In the second part of this paper, we focus on developing the analytic bootstrap for the $\mathcal{N}=1$ super-Ising model. 
At large spin, the spectra of the $\mathcal{N}=1$ super-Ising model approaches the spectra of generalized free fields. Due to the existence of the Majorana fermion, there is operator degeneracy. For example, the operators $\sigma \partial_{\mu_1}\ldots \partial_{\mu_l}\sigma$ and $\psi \gamma_{(\mu_1}\partial_{\mu_2}\ldots \partial_{\mu_l)}\psi$ have the same twist and spin. In interacting theories, these two operators mix and one needs to resolve this mixing effect to obtain accurate conformal data. 
We show that this can be done with the help of superconformal relations, which then allow us to use the non-supersymmetric Lorentzian inversion formula to make predictions for the conformal data of the super-multiplets that belong to the leading Regge trajectories. We then compare the resulting predictions with numerical data from the extremal functional method (EFM) and find good agreement. 

The paper is organized as follows. In section \ref{sec:results}, we push the numerical calculation to very high precision and make new precision determinations of critical exponents for this model. In section \ref{sec:analytic}, we develop the analytic bootstrap, first working out the generalized free theory solution for our setup, then computing the asymptotic behavior of the leading anomalous dimensions, and finally analyzing the non-perturbative predictions from the Lorentzian inversion formula. In section \ref{sec:extremal}, we use the extremal function method (EFM) to extract data about the low-twist spectrum and compare with the analytic predictions.

\section{Numerical Bootstrap}
\label{sec:results}

Our setup for the conformal bootstrap builds on the previous works~\cite{Bashkirov:2013vya, Rong:2018okz,Atanasov:2018kqw} and is identical to the one described in~\cite{Rong:2018okz}, which is an application of the ``long multiplet bootstrap'' idea initiated in \cite{Cornagliotto:2017dup}. In particular, we use the 4 crossing relations arising from the correlators $\<\s\s\s\s\>$ and $\<\s\s\e\e\>$, where $\s$ and $\e$ are the parity-odd and parity-even scalars contained in the leading $\mathcal{B}_-^{(0)}$ multiplet $\Sigma$. The second parity-odd scalar $\s' \in \Sigma'$ is assumed to be isolated. All other $\mathcal{B}_+^{(0)}$ and $\mathcal{B}_-^{(0)}$ multiplets are assumed to have a scaling dimension larger than 3, while all $\mathcal{F}_+^{(1/2)}$ multiplets are assumed to have a scaling dimension larger than $5/2$. All other multiplets are only assumed to satisfy the unitarity bound.

Under these assumptions we have used a Delaunay triangulation search~\cite{Chester:2019ifh} and the convex optimization solver \texttt{sdpb}~\cite{Simmons-Duffin:2015qma,Landry:2019qug} to compute islands in the $\{\ds, \Delta_{\s'}\}$ plane at derivative orders $\Lambda = 35, 43, 51, 59$, improving on the $\Lambda=27$ computation performed in~\cite{Rong:2018okz}. The parameters used for these computations are described in Appendix~\ref{app:parameters}. At $\Lambda=59$ we have computed a total of 92 primal points, which include a fine scan to improve the resolution of the upper-right tip.

\begin{figure} 
  \centering
    \includegraphics[width=.82\textwidth]{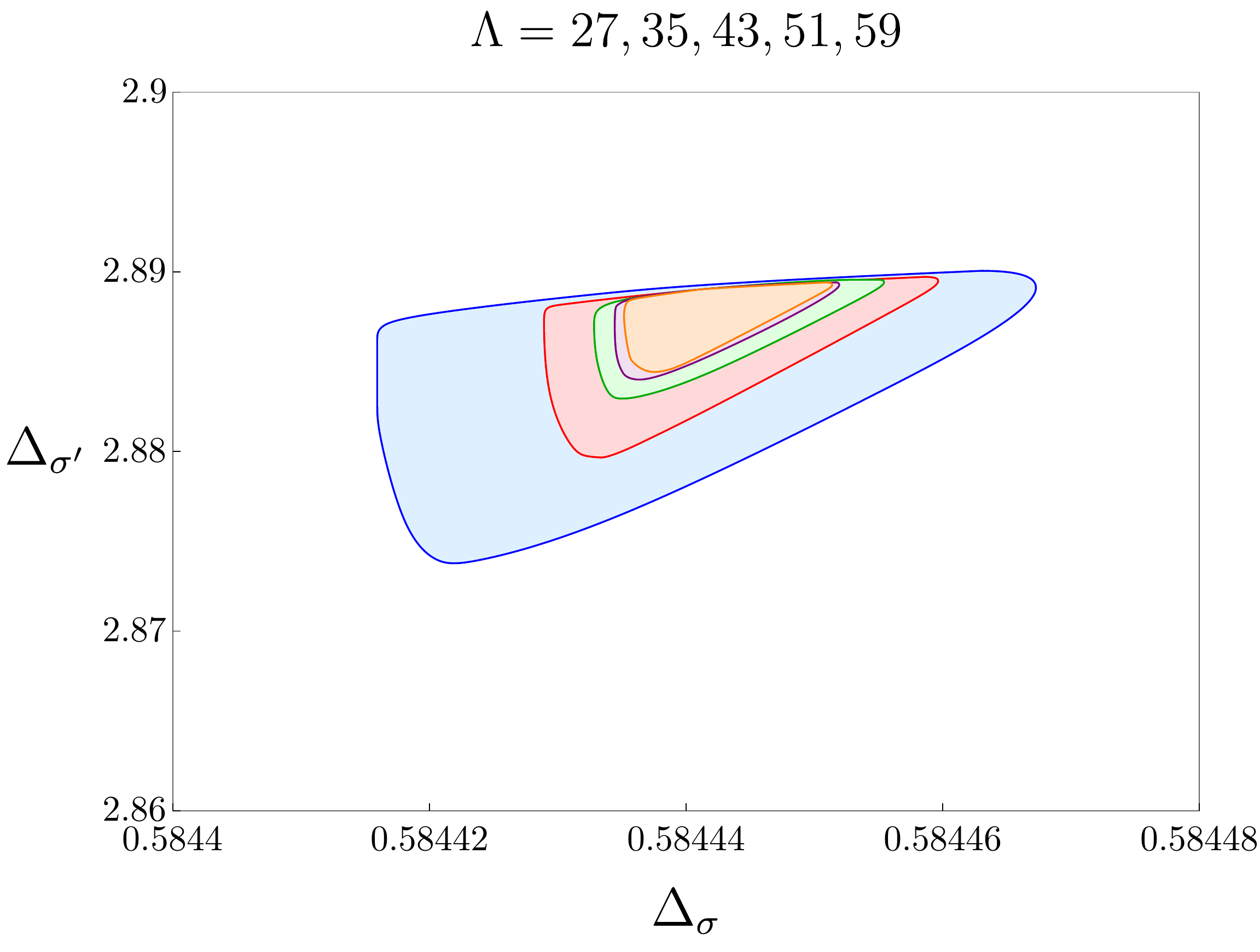}
      \caption{\label{plot:island}Allowed region for the scaling dimensions of the leading parity-odd scalars $\{\s, \s'\}$ in the 3d $\mathcal{N}=1$ super-Ising model.}
\end{figure}

The best determination gives the scaling dimensions
\begin{eqnarray}
\ds &=& 0.5844435(83)\,,\\
\Delta_{\s'} &=& 2.8869(25)\,,
\end{eqnarray}
which translate to the critical exponents
\begin{eqnarray}
\eta_{\sigma} &=& 0.168887(17)\,,\\
\eta_{\psi} &=& 0.168887(17)\,,\\
1/\nu &=& 1.415557(8)\,,\\
\omega &=& 0.8869(25)\,.
\end{eqnarray}
Superconformal symmetry fixes $\lambda_{\e\e\e}/\lambda_{\s\s\e}$ in terms of $\ds$~\cite{Rong:2018okz,Atanasov:2018kqw}, yielding
\begin{eqnarray}
\frac{\lambda_{\e\e\e}}{\lambda_{\s\s\e}} &=& \frac{3(\ds - 1)(3 \ds - 2)}{4 \ds (\ds - 1/2)} \,=\, 1.55775(36) \,=\, \tan(1.00010(11))\,.
\end{eqnarray} 

We will now briefly comment on numerology. We first note that our result is still barely compatible with the possible relation $\lambda_{\e\e\e}/\lambda_{\s\s\e} = \tan(1)$~\cite{Atanasov:2018kqw}, which corresponds to the very upper-right tip of the allowed region. Our value of $\ds$ is also compatible with the possible analytic formula $\ds = (\Gamma(5/24) - 4)/(\Gamma(1/3) - 2) \approx 0.58444186$, which sits in the middle of the allowed region. A sharp goal for future work is to see if we can rule out either of these possibilities. The Inverse Symbolic Calculator \cite{inversesc} is an interesting tool to check whether numerical expressions are compatible with analytic formulas. With higher precision determinations for the scaling dimensions, it will be very interesting to see whether one can identify an analytic expression with a physical meaning behind it. 

Finally, we have computed the minimum and maximum values of the OPE coefficient $\lambda_{\s\s\e}$ at derivative order $\Lambda=51$, by sampling a set of 50 primal $\Lambda = 59$ points across the island. The result is that the OPE coefficient lives in the range
\begin{eqnarray}
\lambda_{\s\s\e} &=& 1.072125(9)\,.
\end{eqnarray}

\section{Analytic Bootstrap}
\label{sec:analytic}

\subsection{Supersymmetric Generalized Free Fields}

Next we will work out the predictions from the analytic bootstrap for $\mathcal{N}=1$ SCFT. To begin we need to discuss the correlators of $\mathcal{N} = 1$ supersymmetric generalized free fields.

As a reminder, the 4-point function of a non-supersymmetric generalized free field $\<\s\s\s\s\>$ can be decomposed into conformal blocks, where exchanged operators of even spin $\ell$ exist at dimension $\Delta_{n,\ell} = 2\ds + 2n + \ell$ with coefficients~\cite{Fitzpatrick:2011dm}\footnote{Here we use the conformal block conventions given by the 1st line of Table I of~\cite{Poland:2018epd}.}
\begin{equation}
\lambda_{\s\s[\s\s]_{n,\ell}}^2 = \frac{(1+(-1)^{\ell}) \times 2^{\ell} (\ds - 1/2)_n^2 (\ds)_{n+\ell}^2}{n! \ell! (n+2\ds -2)_n (\ell + 3/2)_n (n+\ell+2\ds - 3/2)_n (2n+\ell+2\ds - 1)_{\ell}}\,.
\end{equation}
Now we will re-interpret this coefficient in a generalized free theory with $\mathcal{N} = 1$ superconformal symmetry, as describing multiple degenerate contributions. 

In particular, the spin-$\ell$ contributions at $2\ds + 2n + \ell$ can arise from a superconformal primary $[\mathcal{B}^{(\ell)}_+]_{n}$ of dimension $2\ds + 2n + \ell$, as a super-descendant of a multiplet $[\mathcal{B}^{(\ell)}_-]_n$ of dimension $2\ds + 2n + \ell - 1$, or as super-descendants of $[\mathcal{F}_+^{(\ell + 1/2)}]_n$ or $[\mathcal{F}_{-}^{(\ell - 1/2)}]_n$ multiplets of dimension $2\ds + 2n + \ell - 1/2$. Allowing for all types of multiplets, we can write
\begin{eqnarray}
\lambda_{\s\s[\s\s]_{n,\ell}}^2 &=& \lambda_{\s\s[\mathcal{B}^{(\ell)}_+]_{n}}^2 + \lambda_{\s\s Q^2 [\mathcal{B}^{(\ell)}_-]_{n}}^2 + \lambda_{\s\s Q^{-} [\mathcal{F}^{(\ell + 1/2)}_+]_{n}}^2 + \lambda_{\s\s Q^{+} [\mathcal{F}^{(\ell - 1/2)}_-]_{n}}^2\,.
\end{eqnarray}
Here we use a shorthand notation where $Q^{\pm} \mathcal{O}^{(\ell)}_p$ denotes the superdescendant of $\mathcal{O}^{(\ell)}_p$ with spin $\ell \pm 1/2$, dimension $\Delta_{\mathcal{O}} + 1/2$, and parity $\mp p$. Similarly, $Q^2 \mathcal{O}^{(\ell)}_p$ denotes the superdescendant with spin $\ell$, dimension $\Delta_{\mathcal{O}} + 1$, and parity $- p$. 

Now we will consider a second generalized free field $\e$ of dimension $\ds + 1$, which has the GFF coefficients
\begin{equation}
\lambda_{\e\e[\e\e]_{n,\ell}}^2 = \frac{(1+(-1)^{\ell}) \times 2^{\ell} (\ds + 1/2)_n^2 (\ds+1)_{n+\ell}^2}{n! \ell! (n+2\ds )_n (\ell + 3/2)_n (n+\ell+2\ds + 1/2)_n (2n+\ell+2\ds  + 1)_{\ell}}\,.
\end{equation}
Assuming it lives with $\s$ in a supermultiplet, $\e = Q^2 \sigma$, and hence couples to the same operators, we expect a decomposition of the form
\begin{eqnarray}
\lambda_{\e\e[\e\e]_{n-1,\ell}}^2 &=& \lambda_{\e\e[\mathcal{B}^{(\ell)}_+]_{n}}^2 + \lambda_{\e\e Q^2 [\mathcal{B}^{(\ell)}_-]_{n}}^2 + \lambda_{\e\e Q^{-} [\mathcal{F}^{(\ell + 1/2)}_+]_{n}}^2 + \lambda_{\e\e Q^{+} [\mathcal{F}^{(\ell - 1/2)}_-]_{n}}^2\,,\\
&=& \left(\frac{\lambda_{\e\e[\mathcal{B}^{(\ell)}_+]_{n}}}{\lambda_{\s\s[\mathcal{B}^{(\ell)}_+]_{n}}}\right)^2 \lambda_{\s\s[\mathcal{B}^{(\ell)}_+]_{n}}^2 +  \left(\frac{\lambda_{\e\e Q^2 [\mathcal{B}^{(\ell)}_-]_{n}}}{\lambda_{\s\s Q^2[\mathcal{B}^{(\ell)}_-]_{n}}}\right)^2 \lambda_{\s\s Q^2 [\mathcal{B}^{(\ell)}_-]_{n}}^2 \\
&& +  \left(\frac{\lambda_{\e\e Q^{-} [\mathcal{F}^{(\ell + 1/2)}_+]_{n}}}{\lambda_{\s\s Q^{-} [\mathcal{F}^{(\ell + 1/2)}_+]_{n}}}\right)^2  \lambda_{\s\s Q^{-} [\mathcal{F}^{(\ell + 1/2)}_+]_{n}}^2 +   \left(\frac{\lambda_{\e\e Q^{+} [\mathcal{F}^{(\ell - 1/2)}_-]_{n}}}{\lambda_{\s\s Q^{+} [\mathcal{F}^{(\ell - 1/2)}_-]_{n}}}\right)^2  \lambda_{\s\s Q^{+} [\mathcal{F}^{(\ell - 1/2)}_-]_{n}}^2\,.\nonumber
\end{eqnarray}
Here the ratios of OPE coefficients in the second line are fixed by superconformal symmetry and given in Appendix~\ref{app:coefficients}.

We can also consider the mixed correlator $\<\s\s\e\e\>$, which in the $(12)(34)$ channel only contains the identity operator, thus the total non-identity contribution must vanish:
\begin{eqnarray}
0 &=& \lambda_{\s\s[\mathcal{B}^{(\ell)}_+]_{n,\ell}} \lambda_{\e\e[\mathcal{B}^{(\ell)}_+]_{n}} + \lambda_{\s\s Q^2 [\mathcal{B}^{(\ell)}_-]_{n}} \lambda_{\e\e Q^2 [\mathcal{B}^{(\ell)}_-]_{n}} \\
&& + \lambda_{\s\s Q^{-} [\mathcal{F}^{(\ell + 1/2)}_+]_{n}} \lambda_{\e\e Q^{-} [\mathcal{F}^{(\ell + 1/2)}_+]_{n}} + \lambda_{\s\s Q^{+} [\mathcal{F}^{(\ell - 1/2)}_-]_{n}} \lambda_{\e\e Q^{+} [\mathcal{F}^{(\ell - 1/2)}_-]_{n}}\,,\nonumber\\
&=& \left(\frac{\lambda_{\e\e[\mathcal{B}^{(\ell)}_+]_{n}}}{\lambda_{\s\s[\mathcal{B}^{(\ell)}_+]_{n}}}\right) \lambda_{\s\s[\mathcal{B}^{(\ell)}_+]_{n}}^2 +  \left(\frac{\lambda_{\e\e Q^2 [\mathcal{B}^{(\ell)}_-]_{n}}}{\lambda_{\s\s Q^2[\mathcal{B}^{(\ell)}_-]_{n}}}\right) \lambda_{\s\s Q^2 [\mathcal{B}^{(\ell)}_-]_{n}}^2 \\
&& +   \left(\frac{\lambda_{\e\e Q^{-} [\mathcal{F}^{(\ell + 1/2)}_+]_{n}}}{\lambda_{\s\s Q^{-} [\mathcal{F}^{(\ell + 1/2)}_+]_{n}}}\right)  \lambda_{\s\s Q^{-} [\mathcal{F}^{(\ell + 1/2)}_+]_{n}}^2 + \left(\frac{\lambda_{\e\e Q^{+} [\mathcal{F}^{(\ell - 1/2)}_-]_{n}}}{\lambda_{\s\s Q^{+} [\mathcal{F}^{(\ell - 1/2)}_-]_{n}}}\right)  \lambda_{\s\s Q^{+} [\mathcal{F}^{(\ell - 1/2)}_-]_{n}}^2\,.\nonumber
\end{eqnarray}

Finally we can consider the ordering $\<\s\e\s\e\>$, which in the $(12)(34)$ channel yields the GFF coefficient
\begin{eqnarray}
\lambda_{\s\e[\s\e]_{n,\ell}}^2 &=& \frac{2^{\ell} (\ds - 1/2)_n (\ds + 1/2)_n (\ds)_{n+\ell}(\ds+1)_{n+\ell}}{n! \ell! (n+2\ds - 1)_n (\ell + 3/2)_n (n+\ell+2\ds - 1/2)_n (2n+\ell+2\ds)_{\ell}}\,.
\end{eqnarray}
Taking $\ell$ to be even this can then be decomposed as
\begin{align}
\lambda_{\s\e[\s\e]_{n,\ell}}^2 &= \lambda_{\s\e[\mathcal{B}^{(\ell)}_-]_{n+1}}^2 + \lambda_{\s\e Q^2 [\mathcal{B}^{(\ell)}_+]_{n}}^2 \,,\\
&=  \left(\frac{\lambda_{\s\e[\mathcal{B}^{(\ell)}_-]_{n+1}}}{\lambda_{\s\s Q^2 [\mathcal{B}^{(\ell)}_-]_{n+1}}}\right)^2 \lambda_{\s\s Q^2 [\mathcal{B}^{(\ell)}_-]_{n+1}}^2 +  \left(\frac{\lambda_{\s\e Q^2[\mathcal{B}^{(\ell)}_+]_{n}}}{\lambda_{\s\s [\mathcal{B}^{(\ell)}_+]_{n}}}\right)^2\lambda_{\s\s [\mathcal{B}^{(\ell)}_+]_{n}}^2 \,,\\
\lambda_{\s\e[\s\e]_{n,\ell-1}}^2 &= \lambda_{\s\e Q^+[\mathcal{F}^{(\ell - 3/2)}_+]_{n+1}}^2  + \lambda_{\s\e Q^- [\mathcal{F}^{(\ell - 1/2)}_-]_{n}}^2 \,, \\
&= \left( \frac{\lambda_{\s\e Q^+[\mathcal{F}^{(\ell - 3/2)}_+]_{n+1}}}{\lambda_{\s\s Q^-[\mathcal{F}^{(\ell - 3/2)}_+]_{n+1}}} \right)^2 \lambda_{\s\s Q^-[\mathcal{F}^{(\ell - 3/2)}_+]_{n+1}}^2 + \left( \frac{\lambda_{\s\e Q^-[\mathcal{F}^{(\ell - 1/2)}_-]_{n}}}{\lambda_{\s\s Q^+[\mathcal{F}^{(\ell - 1/2)}_-]_{n}}} \right)^2 \lambda_{\s\s Q^+ [\mathcal{F}^{(\ell - 1/2)}_-]_{n}}^2 \,.
\end{align}

\begin{figure}[h] 
  \centering
    \includegraphics[width=1\textwidth]{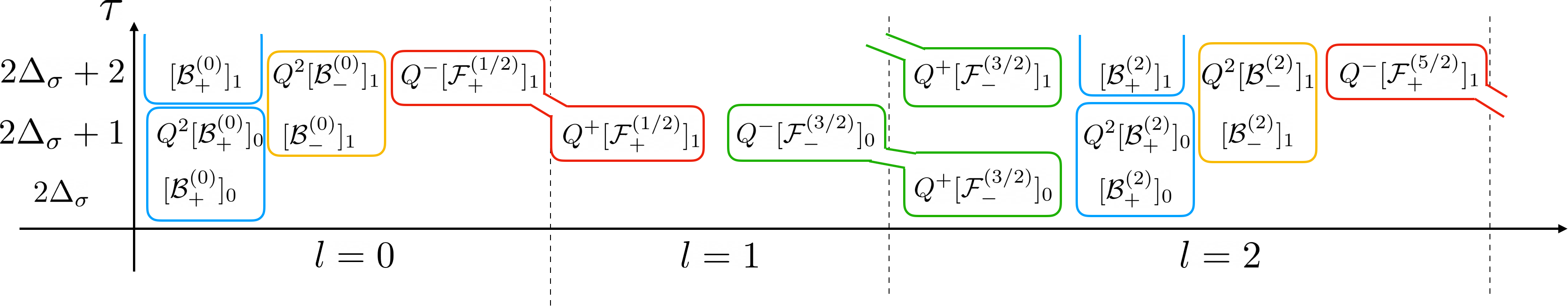}
      \caption{Low-lying operators of SUSY generalized free fields. Circled operators belong to the same multiplet.}\label{SUSYGFF}
\end{figure}

These equations have the solution
\begin{align}\label{eq:GFFBp}
 \lambda_{\s\s[\mathcal{B}^{(\ell)}_+]_{n}}^2 &= \frac{(2 \ds +n-2) (2 \ds +2 n+\ell -1) (4 \ds +2 n+2 \ell -3)}{2 (\ds +n-1) (2 \ds +2 n+2 \ell -1) (4 \ds +4 n+2 \ell -3)} \lambda_{\s\s[\s\s]_{n,\ell}}^2   \,,\\
\label{eq:GFFBm} \lambda_{\s\s Q^2 [\mathcal{B}^{(\ell)}_-]_{n}}^2 &= \frac{n (2 n+2 \ell +1) (2 \ds +2 n+\ell -2)}{2 (\ds +n-1) (2 \ds +2 n+2 \ell -1) (4 \ds +4 n+2 \ell -3)} \lambda_{\s\s[\s\s]_{n,\ell}}^2 \,,\\
\label{eq:GFFFp} \lambda_{\s\s Q^{-} [\mathcal{F}^{(\ell + 1/2)}_+]_{n}}^2 &= \frac{n (\ell +1) (4 \ds +2 n+2 \ell -3)}{2 (2 \ell +1) (\ds +n-1) (2 \ds +2 n+2 \ell -1)} \lambda_{\s\s[\s\s]_{n,\ell}}^2\,,\\
 \label{eq:GFFFm}\lambda_{\s\s Q^{+} [\mathcal{F}^{(\ell - 1/2)}_-]_{n}}^2 &= \frac{\ell  (2 \ds +n-2) (2 n+2 \ell +1)}{2 (2 \ell +1) (\ds +n-1) (2 \ds +2 n+2 \ell -1)} \lambda_{\s\s[\s\s]_{n,\ell}}^2\,.
\end{align}
Note that the coefficients of the $[\mathcal{B}^{(\ell)}_-]_n$ and $[\mathcal{F}^{(\ell+1/2)}_+]_n$ trajectories vanish at $n=0$, so that the leading-twist trajectories correspond to $[\mathcal{B}^{(\ell)}_+]_0$ and $[\mathcal{F}^{(\ell-1/2)}_-]_0$.
The low-lying operators are listed in Figure \ref{SUSYGFF}.

\subsection{Inversion Formula}

As demonstrated in~\cite{Albayrak:2019gnz, Liu:2020tpf, Caron-Huot:2020ouj}, one can use input from the numerical bootstrap along with the Lorentzian inversion formula~\cite{Caron-Huot:2017vep, Simmons-Duffin:2017nub} to make analytic predictions for the spectrum of the theory. 

In the non-supersymmetric case, one defines a generating function whose small $z$ behavior captures the anomalous dimension of the leading-twist trajectory. Assuming this is given by the double-twist tower $[\phi_1 \phi_1]_{0,\ell}$, it will give a leading contribution to the generating function associated to the correlator $\<\phi_1 \phi_1 \phi_2 \phi_2\>$:
\begin{align}\label{eq:generatingNS}
\left(1+ (-1)^{\ell} \right) C_{\phi_1 \phi_1 \phi_2 \phi_2}^t(z,\bar{h}) \approx &\,\, \frac{1}{2^{\ell}} \lambda_{\phi_1 \phi_1 [\phi_1 \phi_1]_{0,\ell}} \lambda_{\phi_2 \phi_2 [\phi_1 \phi_1]_{0,\ell}}  z^{2 h_{1} + \delta h_{[\phi_1 \phi_1]_{0,\ell}}} + \ldots\,, 
\end{align}
where $(h, \bar{h}) \equiv \left(\frac{\Delta - \ell}{2}, \frac{\Delta + \ell}{2}\right) = \left(\frac{\tau}{2}, \frac{\tau}{2} + \ell \right)$ and for convenience we will freely interchange these variables. On the other hand, the inversion formula predicts
\begin{align}\label{eq:Ct}
C^t_{\phi_1 \phi_1 \phi_2 \phi_2}(z,\bar{h}) \approx &\,\, \kappa_{2\bar{h}} \int_0^1 d \bar{z} \frac{1}{\bar{z}^2} k^{0,0}_{\bar{h}}(\bar{z}) \frac{{(z\bar{z})}^{2 h_1}}{[(1-z)(1-\bar{z})]^{h_1 + h_2}} \\
& \times \sum_{\mathcal{O} \in \phi_1 \times \phi_2} 2 \sin^2(\pi(h_{\mathcal{O}} - h_1 - h_2))  (-1)^{\ell_{\mathcal{O}}} \lambda_{\phi_1 \phi_2 \mathcal{O}}^2 g^{h_{21},h_{12}}_{\Delta_{\mathcal{O}},\ell_{\mathcal{O}}}(1-z, 1-\bar{z})\,,\nonumber
\end{align}
where we follow the notation of~\cite{Liu:2020tpf} for the prefactor $\kappa_{2\bar{h}}$ and function $k^{r,s}_{\bar{h}}(\bar{z})$. Similarly, $\<\phi_1 \phi_2 \phi_2 \phi_1\>$ has a generating function:
\begin{align}\label{eq:generatingNS1212}
C^t_{\phi_1 \phi_2 \phi_2 \phi_1}(z,\bar{h}) + (-1)^{\ell} C^u_{\phi_1 \phi_2 \phi_2 \phi_1}(z,\bar{h})  \approx &\,\, \frac{1}{2^{\ell}} \lambda^2_{\phi_1 \phi_2 [\phi_1 \phi_2]_{0,\ell}}  z^{h_{1} + h_2 + \delta h_{[\phi_1 \phi_2]_{0,\ell}}} + \ldots\,, 
\end{align}
with
\begin{align}\label{eq:Ct1221}
C^t_{\phi_1 \phi_2 \phi_2 \phi_1}(z,\bar{h}) \approx &\,\, \kappa_{2\bar{h}} \int_0^1 d \bar{z} \frac{1}{\bar{z}^2} k^{h_{21},h_{12}}_{\bar{h}}(\bar{z}) \frac{{(z\bar{z})}^{h_1 + h_2}}{[(1-z)(1-\bar{z})]^{2h_2}} \\
& \times \sum_{\mathcal{O} \in \phi_1 \times \phi_1, \phi_2 \times \phi_2} 2 \sin(\pi(h_{\mathcal{O}} - 2h_1)) \sin(\pi(h_{\mathcal{O}} - 2h_2)) \nonumber\\
&\times \lambda_{\phi_1 \phi_1 \mathcal{O}} \lambda_{\phi_2 \phi_2 \mathcal{O}} g^{0,0}_{\Delta_{\mathcal{O}},\ell_{\mathcal{O}}}(1-z, 1-\bar{z})\,\nonumber
\end{align}
and
\begin{align}\label{eq:Cu1221}
C^u_{\phi_1 \phi_2 \phi_2 \phi_1}(z,\bar{h}) \approx &\,\, \kappa_{2\bar{h}} \int_0^1 d \bar{z} \frac{1}{\bar{z}^2} k^{h_{21},h_{21}}_{\bar{h}}(\bar{z}) \frac{{(z\bar{z})}^{h_1 + h_2}}{[(1-z)(1-\bar{z})]^{h_1 + h_2}} \\
& \times \sum_{\mathcal{O} \in \phi_1 \times \phi_2} 2 \sin^2(\pi(h_{\mathcal{O}} - h_1 - h_2)) \lambda_{\phi_1 \phi_2 \mathcal{O}}^2 g^{h_{12},h_{12}}_{\Delta_{\mathcal{O}},\ell_{\mathcal{O}}}(1-z, 1-\bar{z})\,.\nonumber
\end{align}
The $\bar{z}$ integrals can be straightforwardly done by expanding the conformal blocks contributing to the integrands in terms of 2d or SL${}_2$ blocks, each of which give a contribution expressible in terms of ${}_4F_3$ hypergeometric functions~\cite{Albayrak:2019gnz, Liu:2020tpf, Sleight:2018epi, Sleight:2018ryu, Cardona:2018dov, Liu:2018jhs, Cardona:2018qrt, Li:2019dix, Li:2019cwm, Li:2020ijq}.

Applying this formalism to an $\mathcal{N} = 1$ SCFT, for $\<\s\s\s\s\>$ and $\<\e\e\e\e\>$, one should write 
\begin{align}\label{eq:generatingssss}
(1+ (-1)^{\ell}) C_{\s\s\s\s}^t(z,\bar{h}) \approx &\,\, \frac{1}{2^{\ell}} \lambda^2_{\s \s [\mathcal{B}^{(\ell)}_+]_0} z^{\ds + \delta h_{[\mathcal{B}^{(\ell)}_+]_0}} \\
& +  \frac{1}{2^{\ell}} \lambda^2_{\s \s Q^+[\mathcal{F}^{(\ell-1/2)}_-]_0}  z^{\ds + \delta h_{[\mathcal{F}^{(\ell-1/2)}_-]_0}} + \ldots\,,\nonumber\\
(1+ (-1)^{\ell}) C_{\e\e\e\e}^t(z,\bar{h}) \approx &\,\, \frac{1}{2^{\ell}} \lambda^2_{\e \e [\mathcal{B}^{(\ell)}_+]_0} z^{\ds + \delta h_{[\mathcal{B}^{(\ell)}_+]_0}} \\
& +  \frac{1}{2^{\ell}} \lambda^2_{\e \e Q^+[\mathcal{F}^{(\ell-1/2)}_-]_0}  z^{\ds + \delta h_{[\mathcal{F}^{(\ell-1/2)}_-]_0}} + \ldots\,,\nonumber
\end{align}
while for $\<\s\s\e\e\>$ one should write
\begin{align}\label{eq:generatingssee}
(1+ (-1)^{\ell}) C_{\s\s\e\e}^t(z,\bar{h}) \approx &\,\, \frac{1}{2^{\ell}} \lambda_{\s \s [\mathcal{B}^{(\ell)}_+]_0}\lambda_{\e \e [\mathcal{B}^{(\ell)}_+]_0} z^{\ds + \delta h_{[\mathcal{B}^{(\ell)}_+]_0}} \\
& +  \frac{1}{2^{\ell}}  \lambda_{\s \s Q^+[\mathcal{F}^{(\ell-1/2)}_-]_0} \lambda_{\e \e Q^+[\mathcal{F}^{(\ell-1/2)}_-]_0} z^{\ds + \delta h_{[\mathcal{F}^{(\ell-1/2)}_-]_0}} + \ldots\,,\nonumber
\end{align}
and for $\<\s\e\e\s\>$ we should write
\begin{align}\label{eq:generatingsees}
C_{\s\e\e\s}^t(z,\bar{h}) + C_{\s\e\e\s}^u(z,\bar{h})  \approx &\,\, \frac{1}{2^{\ell}} \lambda_{\s \e Q^2 [\mathcal{B}^{(\ell)}_+]_0}^2 z^{\ds + \frac12 + \delta h_{[\mathcal{B}^{(\ell)}_+]_0}} \\
& +  \frac{1}{2^{\ell}}  \lambda_{\s \e [\mathcal{B}^{(\ell)}_-]_1}^2 z^{\ds + \frac12 + \delta h_{[\mathcal{B}^{(\ell)}_-]_1}} + \ldots\,, \qquad\qquad (\ell \,\, \text{even}) \nonumber\\
C_{\s\e\e\s}^t(z,\bar{h}) - C_{\s\e\e\s}^u(z,\bar{h})  \approx &\,\, \frac{1}{2^{\ell}} \lambda_{\s \e Q^+ [\mathcal{F}^{(\ell-1/2)}_+]_1}^2 z^{\ds + \frac12 + \delta h_{[\mathcal{F}^{(\ell-1/2)}_+]_1}} \\
& +  \frac{1}{2^{\ell}}  \lambda_{\s \e Q^- [\mathcal{F}^{(\ell+1/2)}_-]_0}^2 z^{\ds + \frac12 + \delta h_{[\mathcal{F}^{(\ell+1/2)}_-]_0}} + \ldots\,.\nonumber \, (\ell \,\, \text{odd})
\end{align}

To a first approximation, we can expand the exponents to linear order in the anomalous dimensions and match $z^{\ds}$ and $z^{\ds} \log(z)$ terms on both sides of the equations. After using the SUSY relations between the OPE coefficients and expanding the exponents to linear order, the $\<\s\s\s\s\>$ correlator gives the conditions
\begin{align}
\label{eq:sssscond1}
\llangle \lambda^2_{\s \s [\s \s]_{0,\ell}} \rrangle\,\,  =&\,\,  \lambda^2_{\s \s [\mathcal{B}^{(\ell)}_+]_0} + \lambda^2_{\s \s Q^+[\mathcal{F}^{(\ell-1/2)}_-]_0} \,,\\
\label{eq:sssscond2}
\llangle \lambda^2_{\s \s [\s \s]_{0,\ell}}  \delta h_{[\s \s]_{0,\ell}} \rrangle\,\, =&\,\, \lambda^2_{\s \s [\mathcal{B}^{(\ell)}_+]_0} \delta h_{[\mathcal{B}^{(\ell)}_+]_0} +  \lambda^2_{\s \s Q^+[\mathcal{F}^{(\ell-1/2)}_-]_0} \delta h_{[\mathcal{F}^{(\ell-1/2)}_-]_0} \,,
\end{align}
while (making use of the analytic expressions for OPE coefficient ratios in Appendix~\ref{app:coefficients}) the $\<\s\s\e\e\>$ correlator gives the conditions
\begin{align}\label{eq:sseecond1}
\llangle \lambda_{\s \s [\s \s]_{0,\ell}} \lambda_{\e \e [\s \s]_{0,\ell}} \rrangle\,\,  =&\,\, \left(  \frac{\delta h_{[\mathcal{B}^{(\ell)}_+]_0}(2\bar{h} - 2 \ds +1)}{\ds(2\ds -1 )}\right) \lambda_{\s \s [\mathcal{B}^{(\ell)}_+]_0}^2   \\
& - \left(\frac{ \delta h_{[\mathcal{F}^{(\ell-1/2)}_-]_0}(2\bar{h} + 2\ds - 3 )}{\ds (2\ds -1)}\right)  \lambda_{\s \s Q^+[\mathcal{F}^{(\ell-1/2)}_-]_0} ^2  \  \,,\nonumber\\
\label{eq:sseecond2}
\llangle \lambda_{\s \s [\s \s]_{0,\ell}} \lambda_{\e \e [\s \s]_{0,\ell}} \delta h_{[\s \s]_{0,\ell}} \rrangle\,\, =&\,\,  \left(  \frac{\delta h_{[\mathcal{B}^{(\ell)}_+]_0}(2\bar{h} - 2 \ds +1)}{\ds(2\ds -1 )}\right) \lambda_{\s \s [\mathcal{B}^{(\ell)}_+]_0}^2 \delta h_{[\mathcal{B}^{(\ell)}_+]_0}   \\
& - \left(\frac{ \delta h_{[\mathcal{F}^{(\ell-1/2)}_-]_0}(2\bar{h} + 2\ds - 3 )}{\ds (2\ds -1)}\right) \lambda_{\s \s Q^+[\mathcal{F}^{(\ell-1/2)}_-]_0}^2 \delta h_{[\mathcal{F}^{(\ell-1/2)}_-]_0}  \,,\nonumber
\end{align}
and similarly the $\<\e\e\e\e\>$ correlator gives
\begin{align}\label{eq:eeeecond1}
\llangle \lambda_{\e\e[\s\s]_{0,\ell}}^2 \rrangle\,\, =&\,\,\left(  \frac{\delta h_{[\mathcal{B}^{(\ell)}_+]_0}(2\bar{h} - 2 \ds +1)}{\ds(2\ds -1 )}\right)^2 \lambda_{\s \s [\mathcal{B}^{(\ell)}_+]_0}^2   \\
& + \left(\frac{ \delta h_{[\mathcal{F}^{(\ell-1/2)}_-]_0}(2\bar{h} + 2\ds - 3 )}{\ds (2\ds -1)}\right)^2  \lambda_{\s \s Q^+[\mathcal{F}^{(\ell-1/2)}_-]_0} ^2  \  \,,\nonumber\\
\label{eq:eeeecond2}
\llangle \lambda_{\e \e [\s \s]_{0,\ell}}^2 \delta h_{[\s \s]_{0,\ell}} \rrangle\,\, =&\,\,  \left(  \frac{\delta h_{[\mathcal{B}^{(\ell)}_+]_0}(2\bar{h} - 2 \ds +1)}{\ds(2\ds -1 )}\right)^2 \lambda_{\s \s [\mathcal{B}^{(\ell)}_+]_0}^2 \delta h_{[\mathcal{B}^{(\ell)}_+]_0}  \\
& + \left(\frac{ \delta h_{[\mathcal{F}^{(\ell-1/2)}_-]_0}(2\bar{h} + 2\ds - 3 )}{\ds (2\ds -1)}\right)^2 \lambda_{\s \s Q^+[\mathcal{F}^{(\ell-1/2)}_-]_0} ^2  \delta h_{[\mathcal{F}^{(\ell-1/2)}_-]_0}  \,.\nonumber
\end{align}
The double bracket notation on the left-hand side of these formulas denotes the averaged values of the analogous non-supersymmetric quantity and can be computed by extracting the coefficients of the $z^{\ds}$ or $z^{\ds} \log z$ terms from the appropriate inversion integrals.  Note that one must take great care with the asymptotic behavior of the combinations $\delta h_{\mathcal{O}} \times \bar{h}$, which we will see shortly are not suppressed at large $\bar{h}$. 

As discussed in~\cite{Simmons-Duffin:2016wlq, Albayrak:2019gnz, Liu:2020tpf, Caron-Huot:2020ouj}, a more refined analysis can be performed by evaluating the equations and their derivatives at finite values of $z$. Concretely, a strategy that works well is to solve~(\ref{eq:generatingssss}) and~(\ref{eq:generatingssee}) for the 2 unknown OPE coefficients $\{\lambda^2_{\s \s [\mathcal{B}^{(\ell)}_+]_0}, \lambda^2_{\s \s Q^+[\mathcal{F}^{(\ell-1/2)}_-]_0} \}$, and then numerically solve the two equations $\{z \frac{d}{dz} \lambda^2_{\s\s  [\mathcal{B}^{(\ell)}_+]_0} = 0, z \frac{d}{dz} \lambda^2_{\s\s Q^+[\mathcal{F}^{(\ell-1/2)}_-]_0} = 0\}$  for the two unknown anomalous dimensions $\{\delta h_{[\mathcal{B}_+^{(\ell)}]_0}, \delta h_{[\mathcal{F}^{(\ell-1/2)}_-]_0}\}$ at some fixed value $z = z_0$.

\subsection{Anomalous Dimensions}
Next let us use our linear approximations to solve for the anomalous dimensions in terms of the OPE coefficients, which will help us extract their asymptotic behavior. We can use~(\ref{eq:sssscond2}) and~(\ref{eq:sseecond1}) to obtain the solution
\begin{align}
 \delta h_{[\mathcal{B}^{(\ell)}_+]_0} =&\,\, \frac{\ds (2\ds-1)}{2(2\bar{h} - 1)} \frac{\llangle \lambda_{\s \s [\s \s]_{0,\ell}} \lambda_{\e \e [\s \s]_{0,\ell}} \rrangle}{\lambda^2_{\s\s{[\mathcal{B}^{(\ell)}_+]_0}}} + \frac{(2\bar{h} + 2\ds -3)}{2(2\bar{h}-1)} \frac{\llangle \lambda^2_{\s \s [\s \s]_{0,\ell}}  \delta h_{[\s \s]_{0,\ell}} \rrangle}{ \lambda_{\s\s{[\mathcal{B}^{(\ell)}_+]_0}}^2}\,,\\
  \delta h_{[\mathcal{F}^{(\ell - 1/2)}_-]_{0}} =&\,\, -\frac{\ds (2\ds-1)}{2(2\bar{h} - 1)} \frac{\llangle \lambda_{\s \s [\s \s]_{0,\ell}} \lambda_{\e \e [\s \s]_{0,\ell}} \rrangle}{\lambda^2_{\s\s{Q^{+} [\mathcal{F}^{(\ell - 1/2)}_-]_{0}}}} + \frac{(2\bar{h} - 2\ds +1)}{2(2\bar{h}-1)} \frac{\llangle \lambda^2_{\s \s [\s \s]_{0,\ell}}  \delta h_{[\s \s]_{0,\ell}} \rrangle}{ \lambda^2_{\s\s{Q^{+} [\mathcal{F}^{(\ell - 1/2)}_-]_{0}}}}\,.
\end{align}

To a first approximation the OPE coefficients can be replaced by the GFF coefficients~(\ref{eq:GFFBp}) and~(\ref{eq:GFFFm}), and in general are expected to have the same asymptotic behavior at large $\bar{h}$, given by
\begin{align}\label{eq:asymptoticOPE}
 \lambda_{\s\s[\mathcal{B}^{(\ell)}_+]_{0}}^2 \sim \lambda_{\s\s Q^{+} [\mathcal{F}^{(\ell - 1/2)}_-]_{0}}^2 \sim \frac{\lambda_{\s\s [\s\s]_{0,\ell}}^2}{2} \sim (1+(-1)^{\ell})\frac{2^{1-\ds}\sqrt{\pi}}{\Gamma(\ds)^2} \frac{1}{2^{\bar{h}}\bar{h}^{\frac{3}{2} - 2\ds}}\,.
\end{align}
This behavior is required from consistent inversion of the identity operator in all correlators.

Then the leading behavior of the anomalous dimensions at large $\bar{h}$ will be given by
\begin{align}
 \delta h_{[\mathcal{B}^{(\ell)}_+]_0} \sim \frac{\ds (2\ds-1)}{2\bar{h}} \frac{\llangle \lambda_{\s \s [\s \s]_{0,\ell}} \lambda_{\e \e [\s \s]_{0,\ell}} \rrangle}{\lambda_{\s\s [\s\s]_{0,\ell}}^2} + \frac{\llangle \lambda^2_{\s \s [\s \s]_{0,\ell}}  \delta h_{[\s \s]_{0,\ell}} \rrangle}{\lambda_{\s\s [\s\s]_{0,\ell}}^2} \,,\\
  \delta h_{[\mathcal{F}^{(\ell-1/2)}_-]_0} \sim -\frac{\ds (2\ds-1)}{2\bar{h}} \frac{\llangle \lambda_{\s \s [\s \s]_{0,\ell}} \lambda_{\e \e [\s \s]_{0,\ell}} \rrangle}{\lambda_{\s\s [\s\s]_{0,\ell}}^2} + \frac{\llangle \lambda^2_{\s \s [\s \s]_{0,\ell}}  \delta h_{[\s \s]_{0,\ell}} \rrangle}{\lambda_{\s\s [\s\s]_{0,\ell}}^2} \,.
\end{align}
The first term gets a dominant contribution from $\s$-exchange and leads to a behavior which falls off like $\sim 1/\bar{h}^{\ds}$, while the second term gets contributions from stress-tensor and $\epsilon$ exchange (the lowest-twist operators in the $\s \times \s$ OPE) which contribute subleading effects $\sim 1/\bar{h}$ and $\sim 1/\bar{h}^{\de}$. 

More precisely, we have 
\begin{align}
\llangle \lambda_{\s \s [\s \s]_{0,\ell}} \lambda_{\e \e [\s \s]_{0,\ell}} \rrangle  \,\,= &\,\, (1+ (-1)^{\ell}) 2^{\ell} C^t_{\s\s\e\e}(z,\bar{h})\bigg|_{z^{\ds}} \\
\approx &\,\, (1+ (-1)^{\ell}) 2^{\bar{h}-\ds} \kappa_{2\bar{h}} \int_0^1 d \bar{z} \frac{1}{\bar{z}^2} k^{0,0}_{\bar{h}}(\bar{z}) \frac{{(z\bar{z})}^{\ds}}{[(1-z)(1-\bar{z})]^{\ds+\frac12}}  \nonumber \\
& \times 2 \cos^2\left(\frac{\pi \ds}{2}\right)  \lambda_{\s \s \e}^2 g^{1/2,-1/2}_{\Delta_{\s},0}(1-z, 1-\bar{z}) \bigg|_{z^{\ds}}\,\nonumber\\
\approx &\,\, (1+ (-1)^{\ell}) 2^{\bar{h}-\ds} \kappa_{2\bar{h}} 2 \cos^2\left(\frac{\pi \ds}{2}\right)  \lambda_{\s \s \e}^2 \nonumber \\
&\,\,\ \times \sum_{p=0}^{\infty} \mathcal{A}_{p,p}^{1/2,-1/2}(h_{\s}, h_{\s}) \frac{\Gamma(2h_{\s}+2p)}{\Gamma(h_{\s}+p+\frac12)^2}\Omega_{\bar{h},h_{\s} + p, 2h_{\s} + 1/2}^{h_{\s},h_{\s},h_{\e},h_{\e}} \,\nonumber\\
\approx &\,\,   (1+ (-1)^{\ell}) \lambda_{\s \s \e}^2 \frac{2^{2-\ds} \sqrt{\pi} \Gamma(\ds)}{\Gamma(\frac{1+\ds}{2})^4} \frac{\bar{h}^{\ds-1/2}}{2^{\bar{h}}} \,, \nonumber
\end{align}
where we used Eq. (3.20) in \cite{Liu:2020tpf} to express the integral in terms of $\Omega$ and $\mathcal{A}$ as defined in that paper. We conclude that we have asymptotic behavior
\begin{align}\label{eq:asymptoticdh}
 \delta h_{[\mathcal{B}^{(\ell)}_+]_0} \sim - \delta h_{[\mathcal{F}^{(\ell-1/2)}_-]_0} \sim \frac{\gamma_{0}}{\bar{h}^{\ds}}\,,
\end{align}
with
\begin{align}\label{eq:coefficientasymptoticdh}
\gamma_{0} =&\,\, (1+ (-1)^{\ell})  \lambda_{\s \s \e}^2 \frac{\ds (2\ds-1) \Gamma(\ds)^3}{4\Gamma(\frac{1+\ds}{2})^4}\,.
\end{align}

\subsection{Extension to Higher-Twist Trajectories}
We can similarly solve for the asymptotic behavior of the higher-twist trajectories. Focusing on the twist $2\ds + 1$ trajectories, the large $\bar{h}$ limit of the GFF coefficients satisfy the relations 
\begin{equation}
	\begin{aligned}
	 \lambda_{\s \e Q^2 [\mathcal{B}^{(\ell)}_+]_0}^2 \sim \lambda_{\s \e [\mathcal{B}^{(\ell)}_-]_1}^2 \sim \lambda_{\s \e Q^- [\mathcal{F}^{(\ell+1/2)}_-]_0}^2 &\sim  \lambda_{\s \e Q^+ [\mathcal{F}^{(\ell-1/2)}_+]_1}^2\\ &\sim \frac{\lambda_{\s \e [\s\e]_{0,\ell}}^2}{2} \sim \frac{\sqrt{\pi}}{\Gamma(\ds)\Gamma(\ds+1)} \frac{\bar{h}^{2\ds-1/2}}{2^{\bar{h}+\ds-1/2}} \,.
	\end{aligned}
\end{equation}
Using the condition
\begin{align}
\llangle \lambda^2_{\s\e[\s\e]_{0,\ell}} \delta h_{[\s\e]_{0,\ell}} \rrangle \,\,=&\,\,  \lambda_{\s \e Q^2 [\mathcal{B}^{(\ell)}_+]_0}^2 \delta h_{[\mathcal{B}_+^{(\ell)}]_0} +  \lambda_{\s \e [\mathcal{B}^{(\ell)}_-]_1}^2 \delta h_{ [\mathcal{B}^{(\ell)}_-]_1} \qquad (\ell \,\, \text{even})
\end{align}
then gives
\begin{align}
\delta h_{ [\mathcal{B}^{(\ell)}_-]_1} \,\,=&\,\, \frac{\llangle \lambda^2_{\s\e[\s\e]_{0,\ell}} \delta h_{[\s\e]_{0,\ell}} \rrangle}{\lambda_{\s \e [\mathcal{B}^{(\ell)}_-]_1}^2} -  \frac{\lambda_{\s \e Q^2 [\mathcal{B}^{(\ell)}_+]_0}^2}{  \lambda_{\s \e [\mathcal{B}^{(\ell)}_-]_1}^2} \delta h_{[\mathcal{B}_+^{(\ell)}]_0} \\
\sim&\,\, 2\frac{\llangle \lambda^2_{\s\e[\s\e]_{0,\ell}} \delta h_{[\s\e]_{0,\ell}} \rrangle}{\lambda_{\s \e [\s\e]_{0,\ell}}^2} -  \delta h_{[\mathcal{B}_+^{(\ell)}]_0} \qquad (\ell \,\, \text{even}). \label{eq:B1}
\end{align}

Similarly,
\begin{align}
\llangle \lambda^2_{\s\e[\s\e]_{0,\ell}} \delta h_{[\s\e]_{0,\ell}} \rrangle \,\,=&\,\, \lambda_{\s \e Q^+ [\mathcal{F}^{(\ell-1/2)}_+]_1}^2 \delta h_{[\mathcal{F}^{(\ell-1/2)}_+]_1} +  \lambda_{\s \e Q^- [\mathcal{F}^{(\ell+1/2)}_-]_0}^2  \delta h_{[\mathcal{F}^{(\ell+1/2)}_-]_0}  \qquad (\ell \,\, \text{odd})
\end{align}
gives
\begin{align}
 \delta h_{[\mathcal{F}^{(\ell-1/2)}_+]_1} \,\,=&\,\, \frac{\llangle \lambda^2_{\s\e[\s\e]_{0,\ell}} \delta h_{[\s\e]_{0,\ell}} \rrangle}{\lambda_{\s \e Q^+ [\mathcal{F}^{(\ell-1/2)}_+]_1}^2} -  \frac{ \lambda_{\s \e Q^- [\mathcal{F}^{(\ell+1/2)}_-]_0}^2  }{  \lambda_{\s \e Q^+ [\mathcal{F}^{(\ell-1/2)}_+]_1}^2} \delta h_{[\mathcal{F}^{(\ell+1/2)}_-]_0}  \\
\sim&\,\, 2\frac{\llangle \lambda^2_{\s\e[\s\e]_{0,\ell}} \delta h_{[\s\e]_{0,\ell}} \rrangle}{\lambda_{\s \e [\s\e]_{0,\ell}}^2} -  \delta h_{[\mathcal{F}^{(\ell+1/2)}_-]_0}  \qquad (\ell \,\, \text{odd})\,. \label{eq:F1}
\end{align}

The inversion integral then gives the dominant contribution
\begin{align}
\llangle \lambda^2_{\s\e[\s\e]_{0,\ell}} \delta h_{[\s\e]_{0,\ell}} \rrangle \,\, =&\,\, 2^{\ell} (C^{t}_{\s\e\e\s}(z,\bar{h}) + (-1)^{\ell} C^{u}_{\s\e\e\s}(z,\bar{h}))\bigg|_{z^{\ds+1/2} \log z} \\
\approx&\,\, (-1)^{\ell} 2^{\bar{h} - \ds - 1/2} \kappa_{2\bar{h}} \int_0^1 d \bar{z} \frac{1}{\bar{z}^2} k^{1/2,1/2}_{\bar{h}}(\bar{z}) \frac{{(z\bar{z})}^{\ds + 1/2}}{[(1-z)(1-\bar{z})]^{\ds+1/2}} \nonumber\\
& \times 2 \cos^2\left(\frac{\pi \ds}{2}\right) \lambda_{\s \s \e}^2 g^{-1/2,-1/2}_{\ds,0}(1-z, 1-\bar{z}) \bigg|_{z^{\ds+1/2} \log z} \nonumber\\
\approx&\,\, (-1)^{\ell} 2^{\bar{h} - \ds - 1/2} \kappa_{2\bar{h}} \times 2 \cos^2\left(\frac{\pi \ds}{2}\right) \lambda_{\s \s \e}^2 \nonumber\\
& \times \sum_{p=0}^{\infty} \mathcal{A}_{p,p}^{-1/2,-1/2}(h_{\s}, h_{\s}) \left(- \frac{\Gamma(2h_{\s}+2p)}{\Gamma(h_{\s}+ p - \frac12)\Gamma(h_{\s}+p +\frac12)}\right) \Omega_{\bar{h}, h_{\s} + p, 2h_{\s} + 1/2}^{h_{\s}, h_{\e}, h_{\s}, h_{\e}} \nonumber\\
\approx&\,\,   -  (-1)^{\ell} \lambda_{\s \s \e}^2  \frac{\Gamma(\ds)^2 \Gamma(\ds+1)}{\Gamma(\frac{\ds-1}{2}) \Gamma(\frac{\ds+1}{2})^3} \frac{\lambda_{\s\e[\s\e]_{0,\ell}}^2}{\bar{h}^{\ds}}   \,,
\end{align}
where we have factored out the asymptotic behavior of the OPE coefficient $\lambda_{\s\e[\s\e]_{0,\ell}}^2$ in order to make the $\bar{h}^{-\ds}$ suppression manifest.
After evaluating~(\ref{eq:B1}) and~(\ref{eq:F1}) and simplifying we obtain the asymptotic behavior
\begin{align}\label{eq:highertwistasymptotics}
\delta h_{ [\mathcal{B}^{(\ell)}_-]_1} \,\,\sim&\,\, -  \delta h_{[\mathcal{F}^{(\ell+1/2)}_+]_1} \sim \frac{\gamma_{1}}{\bar{h}^{\ds}}\,,
\end{align}
with
\begin{align}
\gamma_{1} =  \lambda_{\s \s \e}^2 \frac{(3-4 \ds)\ds \Gamma(\ds)^3}{2\Gamma(\frac{1+\ds}{2})^4} \,.
\end{align}

\subsection{Double-twist Improvement}

By inverting isolated blocks we can approximate the double discontinuities entering Eqs.~(\ref{eq:sssscond1})-(\ref{eq:sseecond1}). On the other hand, Eqs.~(\ref{eq:sseecond2})-(\ref{eq:eeeecond2}) require a resummation of the leading-twist towers inside the inversion integral in order to get a nonzero contribution. Since the OPE coefficients involving $\e$ are enhanced at large $\bar{h}$ due to the SUSY relations, these contributions can be important. 

Let us first consider including the contributions of the $[\mathcal{B}_+^{(\ell)}]_0$ and $Q^+[\mathcal{F}_-^{(\ell-1/2)}]_0$ towers inside the generating function for $\<\s\s\s\s\>$. We will start the sum over spins at an even intermediate value $\ell_0$, anticipating that we will later include operators with spins $< \ell_0$ as isolated contributions. Then we have
\begin{align}
C^{t}_{\s\s\s\s}(z,\bar{h}) \approx &\,\, \kappa_{2\bar{h}} \int_0^1 d \bar{z} \frac{1}{\bar{z}^2} k^{0,0}_{\bar{h}}(\bar{z}) \frac{{(z\bar{z})}^{2 h_{\s}}}{[(1-z)(1-\bar{z})]^{2h_{\s}}} \\
& \times \left(\sum_{\ell = \ell_0}^{\infty} 2 \sin^2(\pi \delta h_{ [\mathcal{B}_+^{(\ell)}]_0})  \lambda_{\s\s[\mathcal{B}_+^{(\ell)}]_0}^2 g^{0,0}_{\Delta_{[\mathcal{B}_+^{(\ell)}]_0},\ell}(1-z, 1-\bar{z})\right.\nonumber\\
& + \left.\sum_{\ell = \ell_0}^{\infty} 2 \sin^2(\pi \delta h_{[\mathcal{F}_-^{(\ell-1/2)}]_0})  \lambda_{\s\s Q^+[\mathcal{F}_-^{(\ell-1/2)}]_0}^2 g^{0,0}_{\Delta_{Q^+[\mathcal{F}_-^{(\ell-1/2)}]_0},\ell}(1-z, 1-\bar{z})\right)\nonumber\\
\approx &\,\, \kappa_{2\bar{h}} \int_0^1 d \bar{z} \frac{1}{\bar{z}^2} k^{0,0}_{\bar{h}}(\bar{z}) \frac{{(z\bar{z})}^{2 h_{\s}}}{[(1-z)(1-\bar{z})]^{2h_{\s}}} g^{0,0}_{2\ds+\ell,\ell}(1-z, 1-\bar{z}) \\
& \times \sum_{\ell = \ell_0}^{\infty} 2 \pi^2 \left(\delta h_{ [\mathcal{B}_+^{(\ell)}]_0}^2 \lambda_{\s\s [\mathcal{B}_+^{(\ell)}]_0}^2 +  \delta h_{[\mathcal{F}_-^{(\ell-1/2)}]_0}^2  \lambda_{\s\s Q^+[\mathcal{F}_-^{(\ell-1/2)}]_0}^2\right) \nonumber \\
\approx &\,\,  \sum_{\ell = \ell_0}^{\infty}  \left( \delta h^2_{[\mathcal{B}^{(\ell)}_+]_0} \frac{\lambda_{\s\s [\mathcal{B}_+^{(\ell)}]_0}^2}{2^{\ell}} +   \delta h^2_{[\mathcal{F}^{(\ell-1/2)}_-]_0}   \frac{\lambda_{\s\s Q^+[\mathcal{F}_-^{(\ell-1/2)}]_0}^2}{2^{\ell}}\right) \nonumber\\
&\times 2 \pi^2 \kappa_{2\bar{h}}  \sum_{p=0}^{\infty} \sum_{q = -p}^{p} \hat{\mathcal{A}}_{p,q}^{0,0}(\ds) z^{\ds} k_{\bar{h}_{\ell}+q}^{0,0}(1-z) \Omega^{\frac{\ds}{2}}_{\bar{h},\ds+p,\ds}\,, \nonumber
\end{align}
similar to Eq.~(3.30) in~\cite{Liu:2020tpf}. We can now compute the asymptotic contribution from the sum using Eqs.~(\ref{eq:asymptoticOPE}) and~(\ref{eq:asymptoticdh}):
\begin{align}
C^{t}_{\s\s\s\s}(z,\bar{h}) 
\approx &\,\,  \sum_{\substack{\bar{h}_\ell = \ell_0 + \ds + \ell\\ \ell=0,2,\ldots}}^{\infty}  2\left(  \frac{\gamma_0^2}{\bar{h}_{\ell}^{2\ds}} \right) \left(\frac{4\sqrt{\pi}}{\Gamma(\ds)^2} \frac{1}{2^{2\bar{h}_{\ell}}\bar{h}_{\ell}^{\frac{3}{2} - 2\ds}}\right) \nonumber\\
&\times 2 \pi^2 \kappa_{2\bar{h}}  \sum_{p=0}^{\infty} \sum_{q = -p}^{p} \hat{\mathcal{A}}_{p,q}^{0,0}(\ds) z^{\ds} k_{\bar{h}_{\ell}+q}^{0,0}(1-z)  \Omega^{\frac{\ds}{2}}_{\bar{h},\ds+p,\ds}\, \nonumber\\
\approx &\,\,  \frac{2 \gamma_0^2}{\Gamma(\ds)^2} \lim_{a\rightarrow 0} \sum_{\substack{\bar{h}_\ell = \ell_0 + \ds + \ell\\ \ell=0,2,\ldots}}^{\infty} \Gamma(-a)^2  S^{0,0}_a(\bar{h}_{\ell}) \nonumber\\
& \times 2 \pi^2 \kappa_{2\bar{h}}  \sum_{p=0}^{\infty} \sum_{q = -p}^{p} \hat{\mathcal{A}}_{p,q}^{0,0}(\ds) z^{\ds} k_{\bar{h}_{\ell}+q}^{0,0}(1-z)  \Omega^{\frac{\ds}{2}}_{\bar{h},\ds+p,\ds}\, \nonumber\\
\approx &\,\,  \frac{4\pi^2 \gamma_0^2}{\Gamma(\ds)^2} \kappa_{2\bar{h}}  \lim_{a\rightarrow 0}  \sum_{p=0}^{\infty} \sum_{q = -p}^{p} \sum_{\substack{\bar{h}^q_\ell = \ell_0 + \ds + q + \ell\\ \ell=0,2,\ldots}}^{\infty} \Gamma(-a)^2  2^{2q} S^{0,0}_a(\bar{h}^q_{\ell})  \hat{\mathcal{A}}_{p,q}^{0,0}(\ds) z^{\ds} k_{\bar{h}^q_{\ell}}^{0,0}(1-z)  \Omega^{\frac{\ds}{2}}_{\bar{h},\ds+p,\ds}\, \nonumber\\
\approx &\,\,  \frac{4\pi^2 \gamma_0^2}{\Gamma(\ds)^2} \kappa_{2\bar{h}}  \sum_{p=0}^{\infty} \sum_{q = -p}^{p}   2^{2q}  \hat{\mathcal{A}}_{p,q}^{0,0}(\ds)  \Omega^{\frac{\ds}{2}}_{\bar{h},\ds+p,\ds} \nonumber\\
&\times \left(\frac14 \log^2 z + A_0(\ell_0 + \ds + q) \log z + B_0(\ell_0 + \ds + q) \right) z^{\ds} \,, \nonumber
\end{align}
where we wrote the coefficient in terms of
\begin{align}
S_a^{r,s}(\bar{h}) \equiv&\,\, \frac{1}{\Gamma(-a-r)\Gamma(-a-s)} \frac{\Gamma(\bar{h}-r)\Gamma(\bar{h}-s)}{\Gamma(2\bar{h}-1)} \frac{\Gamma(\bar{h}-a-1)}{\Gamma(\bar{h}+a+1)}\,,
\end{align}
used Eq.~(6.23) of~\cite{Simmons-Duffin:2016wlq}, and dropped all terms of order $z^{\ds+1}$ or higher. Here the resulting coefficients are given by
\begin{align}
A_0(\bar{h}_0) =&\,\, H_{\bar{h}_0-2} \,,\\
B_0(\bar{h}_0) =&\,\, \frac{\pi^2}{12} + H_{\bar{h}_0-1}\left(H_{\bar{h}_0-1}-\frac{2}{\bar{h}_0-1}\right)+\frac14 \left( \psi^{(1)}\left(\frac{\bar{h}_0}{2}\right) - \psi^{(1)}\left(\frac{\bar{h}_0+1}{2}\right)\right)\,,
\end{align}
where $\psi^{(n)}(x)$ is the polygamma function and $H_{n} = \psi^{(0)}(n+1) + \gamma$ is a harmonic number. 

This yields corrections to the left-hand side of~(\ref{eq:sssscond1}) and~(\ref{eq:sssscond2}) of the form
\begin{align}
\delta \llangle \lambda_{\s\s[\s\s]_{0,\ell}}^2  \rrangle\,\, =&\,\, (1+ (-1)^{\ell}) 2^{\ell} C^t_{\s\s\s\s}\bigg|_{z^{\ds}} \\
\approx&\,\, (1+ (-1)^{\ell}) 2^{\ell} \frac{4\pi^2 \gamma_0^2}{\Gamma(\ds)^2} \kappa_{2\bar{h}} \nonumber\\
& \sum_{p=0}^{\infty} \sum_{q = -p}^{p}   2^{2q}  \hat{\mathcal{A}}_{p,q}^{0,0}(\ds)  \Omega^{\frac{\ds}{2}}_{\bar{h},\ds+p,\ds} B_0(\ell_0 + \ds + q), \nonumber
\end{align}
\begin{align}
\delta \llangle \lambda_{\s\s[\s\s]_{0,\ell}}^2 \delta h_{[\s\s]_{0,\ell}} \rrangle\,\, =&\,\, (1+ (-1)^{\ell}) 2^{\ell} C^t_{\s\s\s\s}\bigg|_{z^{\ds} \log(z)} \\
\approx&\,\, (1+ (-1)^{\ell}) 2^{\ell} \frac{4\pi^2 \gamma_0^2}{\Gamma(\ds)^2} \kappa_{2\bar{h}}  \nonumber\\
& \sum_{p=0}^{\infty} \sum_{q = -p}^{p}   2^{2q}  \hat{\mathcal{A}}_{p,q}^{0,0}(\ds)  \Omega^{\frac{\ds}{2}}_{\bar{h},\ds+p,\ds} A_0(\ell_0 + \ds + q) \nonumber\,.
\end{align}

For $\<\e\e\e\e\>$, a similar computation gives
\begin{align}
C^{t}_{\e\e\e\e}(z,\bar{h}) \approx &\,\, \kappa_{2\bar{h}} \int_0^1 d \bar{z} \frac{1}{\bar{z}^2} k^{0,0}_{\bar{h}}(\bar{z}) \frac{{(z\bar{z})}^{2 h_{\e}}}{[(1-z)(1-\bar{z})]^{2h_{\e}}} \\
& \times \left(\sum_{\ell = \ell_0}^{\infty} 2 \sin^2(\pi(2h_{\s} + \delta h_{ [\mathcal{B}_+^{(\ell)}]_0} - 2 h_{\e}))  \lambda_{\e\e [\mathcal{B}_+^{(\ell)}]_0}^2 g^{0,0}_{\Delta_{[\mathcal{B}_+^{(\ell)}]_0},\ell}(1-z, 1-\bar{z})\right.\nonumber\\
& + \left.\sum_{\ell = \ell_0}^{\infty} 2 \sin^2(\pi(2h_{\s} + \delta h_{[\mathcal{F}_-^{(\ell-1/2)}]_0} - 2 h_{\e}))  \lambda_{\e\e Q^+[\mathcal{F}_-^{(\ell-1/2)}]_0}^2 g^{0,0}_{\Delta_{Q^+[\mathcal{F}_-^{(\ell-1/2)}]_0},\ell}(1-z, 1-\bar{z})\right)\nonumber\\
\approx &\,\, \kappa_{2\bar{h}} \int_0^1 d \bar{z} \frac{1}{\bar{z}^2} k^{0,0}_{\bar{h}}(\bar{z}) \frac{{(z\bar{z})}^{2 h_{\e}}}{[(1-z)(1-\bar{z})]^{2h_{\e}}} \\
& \times \sum_{\ell = \ell_0}^{\infty} 2 \pi^2 \left(\delta h_{ [\mathcal{B}_+^{(\ell)}]_0}^2 \lambda_{\e\e [\mathcal{B}_+^{(\ell)}]_0}^2 +  \delta h_{[\mathcal{F}_-^{(\ell-1/2)}]_0}^2  \lambda_{\e\e Q^+[\mathcal{F}_-^{(\ell-1/2)}]_0}^2\right) g^{0,0}_{2\ds+\ell,\ell}(1-z, 1-\bar{z})\nonumber \\
\approx &\,\,  \sum_{\ell = \ell_0}^{\infty}  \left(\left(  \frac{\delta h^2_{[\mathcal{B}^{(\ell)}_+]_0}(2\bar{h}_{\ell} - 2\ds  +1)}{\ds(2\ds -1 )}\right)^2 \frac{\lambda_{\s\s [\mathcal{B}_+^{(\ell)}]_0}^2}{2^{\ell}} \right. \nonumber\\
& \qquad \left. +   \left(\frac{ \delta h^2_{[\mathcal{F}^{(\ell-1/2)}_-]_0}(2\bar{h}_{\ell} + 2\ds - 3 )}{\ds (2\ds -1)}\right)^2   \frac{\lambda_{\s\s Q^+[\mathcal{F}_-^{(\ell-1/2)}]_0}^2}{2^{\ell}}\right) \nonumber\\
&\times 2 \pi^2 \kappa_{2\bar{h}}  \sum_{p=0}^{\infty} \sum_{q = -p}^{p} \hat{\mathcal{A}}_{p,q}^{0,0}(\ds) \frac{z^{\de} k_{\bar{h}_{\ell}+q}^{0,0}(1-z)}{(1-z)^{\de}}  \Omega^{\frac{\de}{2}}_{\bar{h},\ds+p,\de}\,. \nonumber
\end{align}
Using Eqs.~(\ref{eq:asymptoticOPE}) and~(\ref{eq:asymptoticdh}) we find a contribution of the form
\begin{align}
C^{t}_{\e\e\e\e}(z,\bar{h}) 
\approx &\,\,  \sum_{\ell = \ell_0}^{\infty}  8\left(  \frac{\gamma_0^4  \bar{h}_{\ell}^{2-4\ds}}{\ds^2(2\ds -1 )^2}\right) \left(\frac{4\sqrt{\pi}}{\Gamma(\ds)^2} \frac{1}{2^{2\bar{h}_{\ell}}\bar{h}_{\ell}^{\frac{3}{2} - 2\ds}}\right) \nonumber\\
&\times 2 \pi^2 \kappa_{2\bar{h}}  \sum_{p=0}^{\infty} \sum_{q = -p}^{p} \hat{\mathcal{A}}_{p,q}^{0,0}(\ds) \frac{z^{\de} k_{\bar{h}_{\ell}+q}^{0,0}(1-z)}{(1-z)^{\de}}  \Omega^{\frac{\de}{2}}_{\bar{h},\ds+p,\de}\,. \nonumber
\end{align}
The coefficient inside the sum is proportional to $S^{0,0}_{\ds-1}(\bar{h}_{\ell})$ so this will contribute terms of order $z^{\de + \ds - 1} = z^{2\ds}$, as well as terms of order $z^{\de}$ and $z^{\de} \log(z)$ to the generating function, giving no contribution to the twist $2\ds$ towers but requiring the appearance of operators with twist approaching $4\ds$ and $2\de$. 

Finally, let us consider the generating function of the mixed correlator $\<\s\s\e\e\>$, which receives contributions from the operators of asymptotic twist $2\ds + 1$:
\begin{align}
C^{t}_{\s\s\e\e}(z,\bar{h}) \approx &\,\, \kappa_{2\bar{h}} \int_0^1 d \bar{z} \frac{1}{\bar{z}^2} k^{0,0}_{\bar{h}}(\bar{z}) \frac{{(z\bar{z})}^{2 h_{\s}}}{[(1-z)(1-\bar{z})]^{h_{\s}+ h_{\e}}} \\
& \times \left[\sum_{\ell = \ell_0, \ell_0 + 2, \ldots}^{\infty} \left(2 \sin^2(\pi(\delta h_{ [\mathcal{B}_+^{(\ell)}]_0}))  (-1)^{\ell} \lambda_{\s\e Q^2[\mathcal{B}_+^{(\ell)}]_0}^2 g^{\frac12,-\frac12}_{\Delta_{Q^2[\mathcal{B}_+^{(\ell)}]_0},\ell}(1-z, 1-\bar{z})\right.\right. \nonumber\\
& \left.\left. \qquad\qquad\qquad + 2 \sin^2(\pi(\delta h_{ [\mathcal{B}_-^{(\ell)}]_1}))  (-1)^{\ell} \lambda_{\s\e [\mathcal{B}_-^{(\ell)}]_1}^2  g^{\frac12,-\frac12}_{\Delta_{[\mathcal{B}_-^{(\ell)}]_1},\ell}(1-z, 1-\bar{z}) \right) \right.\nonumber\\
& \left.+ \sum_{\ell = \ell_0 + 1, \ell_0 + 3, \ldots}^{\infty} \left(2 \sin^2(\pi(\delta h_{ [\mathcal{F}_-^{(\ell+\frac12)}]_0}))  (-1)^{\ell} \lambda_{\s\e Q^-[\mathcal{F}_-^{(\ell+\frac12)}]_0}^2 g^{\frac12,-\frac12}_{\Delta_{Q^-[\mathcal{F}_-^{(\ell+\frac12)}]_0},\ell}(1-z, 1-\bar{z}) \right. \right. \nonumber\\
& \left.\left.\qquad\qquad\qquad + 2 \sin^2(\pi(\delta h_{ [\mathcal{F}_+^{(\ell-\frac12)}]_1})) (-1)^{\ell} \lambda_{\s\e Q^+[\mathcal{F}_+^{(\ell-\frac12)}]_1}^2 g^{\frac12,-\frac12}_{\Delta_{Q^+[\mathcal{F}_+^{(\ell-\frac12)}]_1},\ell}(1-z, 1-\bar{z}) \right) \right] \nonumber\\
\approx &\,\, \kappa_{2\bar{h}} \int_0^1 d \bar{z} \frac{1}{\bar{z}^2} k^{0,0}_{\bar{h}}(\bar{z}) \frac{{(z\bar{z})}^{2 h_{\s}}}{[(1-z)(1-\bar{z})]^{h_{\s}+ h_{\e}}} \nonumber\\
& \times \sum_{\ell = \ell_0}^{\infty} 2\pi^2 \left[\left(\frac{\gamma_0}{\bar{h}_{\ell}^{\ds}}\right)^2 + \left(\frac{\gamma_1}{\bar{h}_{\ell}^{\ds}}\right)^2 \right] \frac{\lambda_{\s\e[\s\e]_{0,\ell}}^2}{2} (-1)^{\ell} g^{\frac12,-\frac12}_{2\ds+1+\ell,\ell}(1-z, 1-\bar{z}) \nonumber\\
\approx &\,\, \sum_{\ell = \ell_0}^{\infty} \left[\left(\frac{\gamma_0}{\bar{h}_{\ell}^{\ds}}\right)^2 + \left(\frac{\gamma_1}{\bar{h}_{\ell}^{\ds}}\right)^2 \right] \left(\frac{2\sqrt{\pi}}{\Gamma(\ds)\Gamma(\ds+1)}\frac{\bar{h}_{\ell}^{2\ds-1/2}}{2^{2\bar{h}_{\ell}}} \right) \nonumber\\
& \times 2\pi^2 \kappa_{2\bar{h}} \sum_{p=0}^{\infty} \sum_{q=-p}^{p} \hat{\mathcal{A}}^{\frac12,-\frac12}_{p,q}(\ds+1/2) z^{\ds} k_{\bar{h}_{\ell}+q}^{\frac12,-\frac12}(1-z) \Omega^{\frac{\ds}{2},\frac{\ds}{2},\frac{\de}{2},\frac{\de}{2}}_{\bar{h},\ds+\frac12+p,\ds+\frac12} \nonumber\\
\approx &\,\, \frac{\left(\gamma_0^2 + \gamma_1^2 \right)}{2\Gamma(\ds)\Gamma(\ds+1)} \sum_{p=0}^{\infty} \sum_{q=-p}^{p} \lim_{a\rightarrow -1/2} \sum_{\substack{\bar{h}_{\ell}^{q} = \ell_0+\ds+1/2+q+\ell\\\ell=0,1,2,\ldots}}^{\infty} 2^{2q} \Gamma(-1/2-a) S_a^{\frac12,-\frac12}(\bar{h}^q_{\ell}) k_{\bar{h}_{\ell}^q}^{\frac12,-\frac12}(1-z) \nonumber\\
& \times 2\pi^2 \kappa_{2\bar{h}}  \hat{\mathcal{A}}^{\frac12,-\frac12}_{p,q}(\ds+1/2) z^{\ds}  \Omega^{\frac{\ds}{2},\frac{\ds}{2},\frac{\de}{2},\frac{\de}{2}}_{\bar{h},\ds+\frac12+p,\ds+\frac12} \nonumber\\
\approx &\,\, - \frac{\pi^2 \left(\gamma_0^2 + \gamma_1^2 \right)}{\Gamma(\ds)\Gamma(\ds+1)} \sum_{p=0}^{\infty} \sum_{q=-p}^{p} 2^{2q} \hat{\mathcal{A}}^{\frac12,-\frac12}_{p,q}(\ds+1/2)  \kappa_{2\bar{h}}  \Omega^{\frac{\ds}{2},\frac{\ds}{2},\frac{\de}{2},\frac{\de}{2}}_{\bar{h},\ds+\frac12+p,\ds+\frac12} \nonumber\\ 
& \times z^{\ds} \left(\log(z) + 2 H_{\ell_0+\ds+q-1}\right) \,. \nonumber
\end{align}
In the last line we extracted the leading $\log(z)$ and regular terms using Eq.~(4.47) of~\cite{Simmons-Duffin:2016wlq}. Note that in the first line we assumed $\ell_0$ was even for concreteness, but the final formula is valid for both even and odd $\ell_0$. 

Thus, the leading term on the left-hand side of~(\ref{eq:sseecond2}) is given by
\begin{align}
\llangle \lambda_{\s\s[\s\s]_{0,\ell}} \lambda_{\e\e[\s\s]_{0,\ell}} \delta h_{[\s\s]_{0,\ell}}\rrangle \,\, = &\,\, (1+(-1)^{\ell}) 2^{\ell} C^{t}_{\s\s\e\e} \bigg|_{z^{\ds} \log(z)} \\
\approx &\,\, -(1+(-1)^{\ell}) 2^{\ell}\frac{\pi^2 \left(\gamma_0^2 + \gamma_1^2 \right)}{\Gamma(\ds)\Gamma(\ds+1)} \kappa_{2\bar{h}} \nonumber\\
& \times \sum_{p=0}^{\infty} \sum_{q=-p}^{p} 2^{2q} \hat{\mathcal{A}}^{\frac12,-\frac12}_{p,q}(\ds+1/2) \Omega^{\frac{\ds}{2},\frac{\ds}{2},\frac{\de}{2},\frac{\de}{2}}_{\bar{h},\ds+\frac12+p,\ds+\frac12} \,,\nonumber
\end{align}
while the correction to the left-hand side of~(\ref{eq:sseecond1}) is given by
\begin{align}
\delta \llangle \lambda_{\s\s[\s\s]_{0,\ell}} \lambda_{\e\e[\s\s]_{0,\ell}} \rrangle \,\, = &\,\, (1+(-1)^{\ell}) 2^{\ell} C^{t}_{\s\s\e\e} \bigg|_{z^{\ds}} \\
\approx &\,\, -(1+(-1)^{\ell}) 2^{\ell} \frac{2\pi^2 \left(\gamma_0^2 + \gamma_1^2 \right)}{\Gamma(\ds)\Gamma(\ds+1)}  \kappa_{2\bar{h}} \nonumber\\
& \times \sum_{p=0}^{\infty} \sum_{q=-p}^{p} 2^{2q} \hat{\mathcal{A}}^{\frac12,-\frac12}_{p,q}(\ds+1/2) \Omega^{\frac{\ds}{2},\frac{\ds}{2},\frac{\de}{2},\frac{\de}{2}}_{\bar{h},\ds+\frac12+p,\ds+\frac12} H_{\ell_0+\ds+q-1} \,.\nonumber
\end{align}

\section{Extremal Spectrum}
\label{sec:extremal}

Using the extremal functional method~\cite{Poland:2010wg, El-Showk:2012vjm, Simmons-Duffin:2016wlq} we can obtain numerical estimates of the higher spectrum of the theory and compare these to predictions from the analytic bootstrap. We have applied this method using the script \texttt{spectrum.py}~\cite{Komargodski:2016auf, spectrum} to extract extremal spectra correponding to the minima and maxima of the OPE coefficient $\lambda_{\s\s\e}$ described in the previous section. Overall, the resulting spectra are quite unstable compared to, e.g., the Ising~\cite{Simmons-Duffin:2016wlq} and O(2) models~\cite{Liu:2020tpf}. We attribute this to the significant amount of operator mixing in the theory which is difficult to numerically disentangle. However, we have been able to extract a few seemingly robust features from these spectra, described below.

\subsection{Leading Scalars and Low Spin}

\begin{figure}[htbp]
  \centering
  \includegraphics[width=0.85\textwidth]{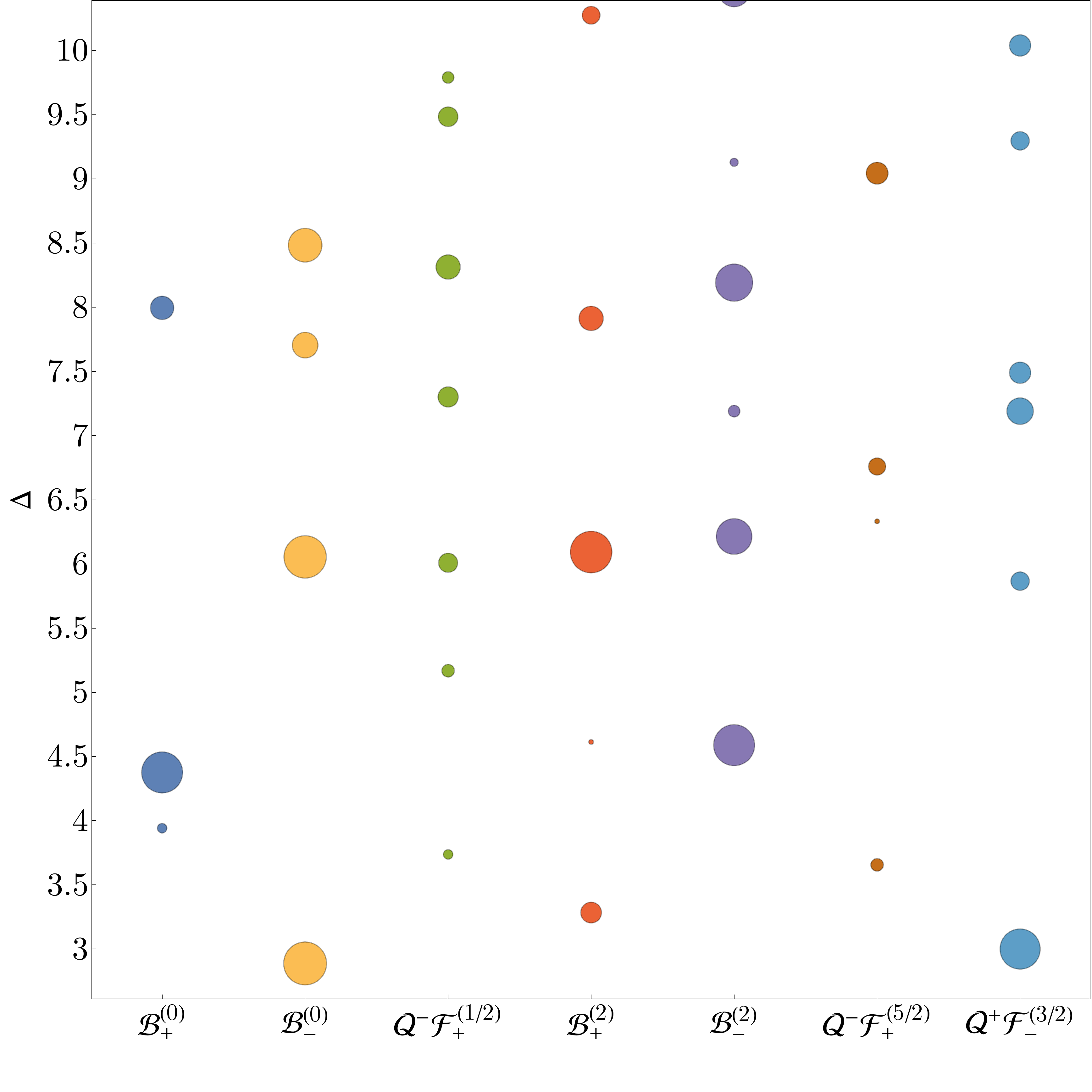}
  \caption{\label{plot:lowspin} Scaling dimensions of low-spin operators appearing in the extremal spectra, where the size of each bubble is proportional to the number of extremal spectra in which the operator is found.}
\end{figure}

In Fig.~\ref{plot:lowspin} we show the locations of low-spin operators in the extremal spectra, where each bubble denotes a collection of operators appearing near that scaling dimension across the extremal spectra and the size of the bubble is proportional to the number of extremal spectra in which the operator is found. 

In the $\mathcal{B}^{(\ell)}_{+}$ sectors we typically find a scalar operator at dimension $4.38(1)$ and a spin-2 operator at dimension $3.28(1)$. In the $\mathcal{B}^{(\ell)}_{-}$ sectors we consistently find the next scalar operator after $\Sigma'$ around dimension $6.0(1)$ and the leading spin-2 operator at dimension $4.58(1)$. In the $\mathcal{F}^{(\ell-1/2)}_{+}$ sectors the spectra show significant fluctuations, with the leading $\ell=1$ multiplet typically containing a scalar component with dimension $\sim 5 - 9$. The extremal spectra are also consistent with a $\mathcal{F}_{-}^{(3/2)}$ stress-tensor multiplet at $\Delta = 5/2$, as expected for a local CFT.

\subsection{Leading-Twist Trajectories}

Many of the extremal spectra also show clear trajectories in the $\mathcal{F}^{(\ell)}_-$ and $\mathcal{B}^{(\ell)}_+$ sectors, where the component appearing in the $\s \times \s$ OPE has twist near $2\ds$. These trajectories can be compared with the predictions from the Lorentzian inversion formula discussed in the previous section. While some of the extremal spectra only have partial trajectories and behave somewhat erratically, others are nearly complete and show a smooth behavior. 

\begin{figure}[tbp]
  \centering
  \includegraphics[width=.85\textwidth]{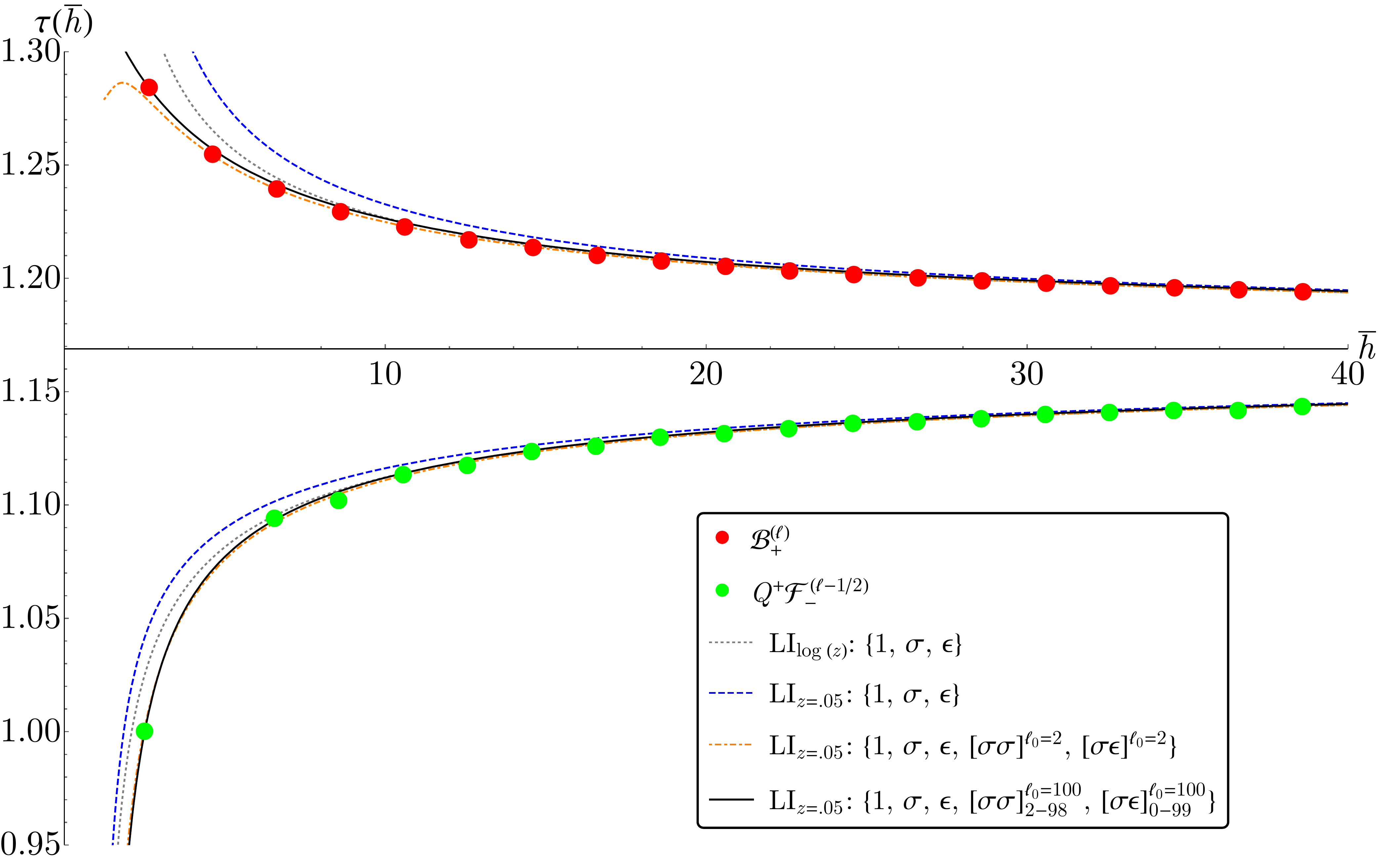}
  \caption{\label{plot:leadingtwist} Leading twist trajectories in the $\mathcal{B}^{(\ell)}_{+}$ and $\mathcal{F}^{(\ell)}_{-}$ sectors from the extremal functional method compared with analytic predictions from the Lorentzian inversion formula after inputting $t$-channel exchange of various sets of operators. The shown extremal spectrum was computed at $\Lambda = 51$ by minimizing the OPE coefficient $\lambda_{\s \s \e}$ at the point $\{\Delta_{\s}, \Delta_{\s'}\} = \{{0.5844353559, 2.888214659}\}$. The horizontal axis is located at $\tau = 2 \ds$.}
\end{figure}

We show an example of such a trajectory in Fig.~\ref{plot:leadingtwist},\footnote{This particular trajectory is missing a spin-4 $Q^+[\mathcal{F}_-^{(7/2)}]_0$ operator, which may induce some error in the nearby spectrum, but is otherwise the cleanest trajectory we have obtained.} where one can see an excellent agreement with the Lorentzian inversion formula after resolving the mixing effects as described in the previous section. In this plot we show the result from including the exchanged operators $\{\mathbb{1}, \s, \e\}$ using the leading $\log(z)$ expansion, along with the results from matching at the finite value $z=.05$ with various choices of exchanged operators. All computations are performed using dimensional reduction and truncating the sum over $p$ at order $p_{\text{max}} = 4$. 

\begin{figure}[tbp]
  \centering
  \includegraphics[width=.85\textwidth]{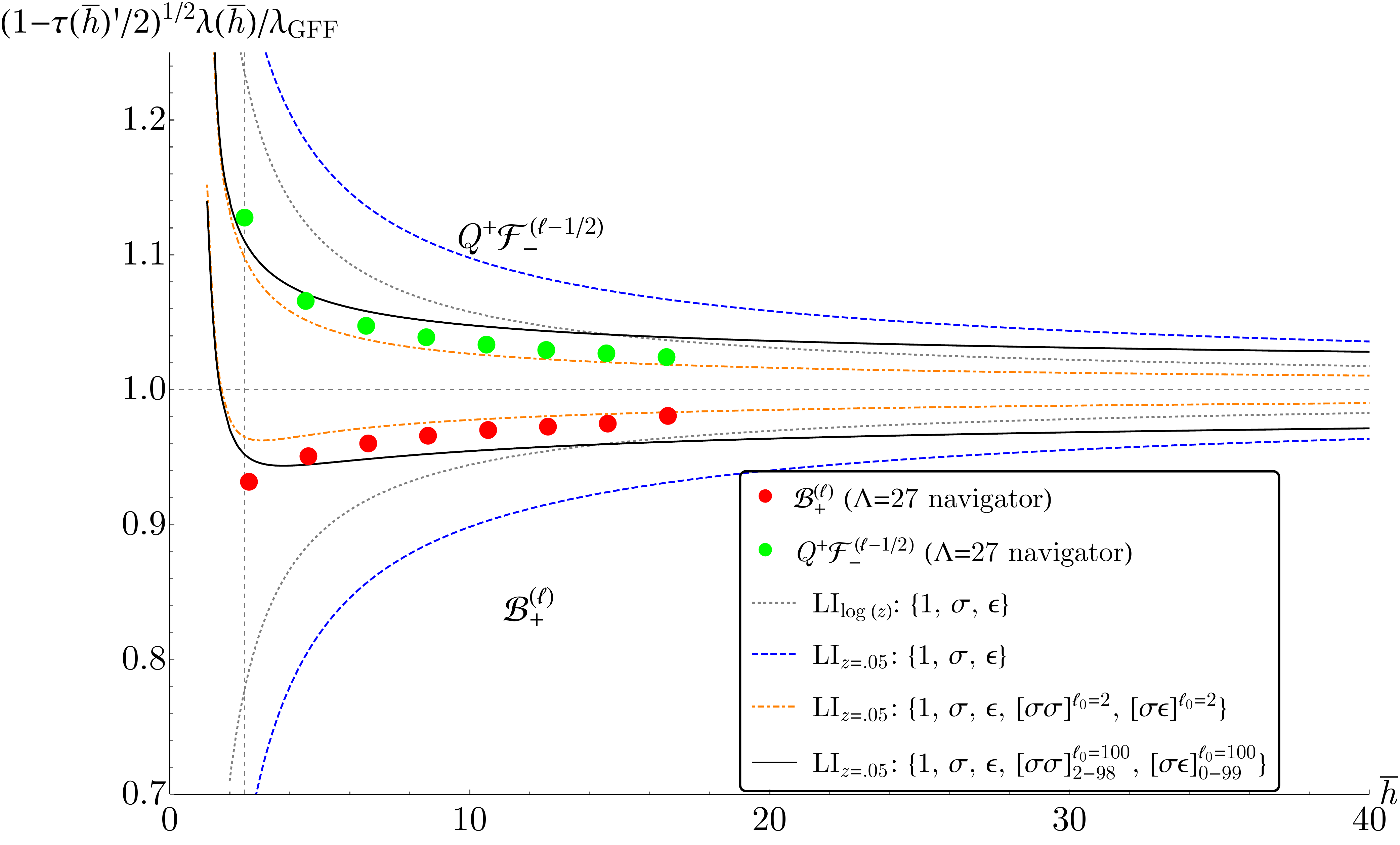}
  \caption{\label{plot:leadingOPE} Leading twist OPE coefficients in the $\mathcal{B}^{(\ell)}_{+}$ and $\mathcal{F}^{(\ell)}_{-}$ sectors from the Lorentzian inversion formula after inputting $t$-channel exchange of various sets of operators. The vertical dashed line is at the stress-tensor location $\bar{h}=5/2$. The curves are compared with extremal spectrum data from the minimum of the navigator function at $\Lambda = 27$.}
\end{figure}

The notation $[\s\s]^{\ell_0=100}_{2-98}$ and $[\s\e]^{\ell_0=100}_{0-99}$ means we have included the double-twist improvements described in the previous section with the specified value of $\ell_0$, along with isolated contributions in the trajectories of asymptotic twist $2\ds$ and $2\ds+1$ for spins below 100, obtained from the analytic solution from the previous set $\{\mathbb{1},\s,\e, [\s\s]_{\ell_0=2}, [\s\e]_{\ell_0=2}\}$. The isolated contributions for the operators in the $[\mathcal{B}_-^{(\ell)}]_1$ and $[\mathcal{F}_+^{(\ell-1/2)}]_1$ multiplets were approximated using the asymptotic formula~(\ref{eq:highertwistasymptotics}) for their scaling dimensions, with the exception of $\Sigma' = [\mathcal{B}_-^{(0)}]_1$ where we used the same value as was used to compute the extremal spectrum, $\Delta_{\s'} = 2.888214659$. For their OPE coefficients we used the approximations
\begin{align}
\lambda_{\s\e[\mathcal{B}_-^{(\ell)}]_1} &\approx d_2[2\ds+\ell+1+2\delta h_{[\mathcal{B}_-^{(\ell)}]_1},\ell]^{1/2} \lambda^{\text{GFF}}_{\s\s Q^2[\mathcal{B}_-^{(\ell)}]_1} \,,\\
\lambda_{\s\e Q^+[\mathcal{F}_+^{(\ell-1/2)}]_1} &\approx f_2[2\ds+\ell+1+2\delta h_{[\mathcal{F}_+^{(\ell-1/2)}]_1},\ell-1]^{1/2} \lambda^{\text{GFF}}_{\s\s Q^-[\mathcal{F}_+^{(\ell-1/2)}]_1}\,.
\end{align}

These approximations can be improved in future analyses, but already one can see excellent agreement. We also see the importance of resumming the leading twist trajectories in order to obtain good predictions at low $\bar{h}$. In the most precise spectrum where we have included isolated contributions for spins up to $\ell_0 = 100$, we see excellent agreement with the existence of a twist $\tau=1$ stress tensor when the $Q^+ [\mathcal{F}^{(\ell-1/2)}_{-}]_0$ trajectory is extrapolated down to $\bar{h} = 5/2$, where the inversion formula gives $\tau \sim 0.9996$. 

The corresponding predictions for the OPE coefficients (normalized to the SUSY generalized free values) are shown in Fig.~\ref{plot:leadingOPE}. In order to relate the coefficients $\lambda^2(\bar{h})$ as a function of $\bar{h}$ to the coefficients at physical spins $\ell$, we must numerically solve the equation $\ell = \bar{h} - \frac{\tau(\bar{h})}{2}$ for integer $\ell$ and include the Jacobian factor $\lambda^2(\ell) = \left(1-\frac{\tau'(\bar{h})}{2}\right)^{-1} \lambda^2(\bar{h})$. We can see that the OPE coefficients are more sensitive than the spectrum to the precise operators included, particularly at low $\bar{h}$. Converting the OPE coefficient of the stress tensor into the central charge, our most precise spectrum gives the estimate $C_T / C_T^{\text{free}} \approx 1.737$, close to the previous estimates $C_T / C_T^{\text{free}} \approx 1.684$ from the numerical bootstrap~\cite{Rong:2018okz} and in excellent agreement with $C_T / C_T^{\text{free}} \approx 1.73$ from a 2-sided Pad\'e${}_{[1,1]}$ approximation applied to the $\epsilon$-expansion~\cite{Fei:2016sgs}. We show the corresponding CFT data in table~\ref{tab:spectrum}, which represents our current best analytic computation of the spectrum.

\begin{table}
\begin{center}
\begin{tabular}{|c|c|c|c|c|}
\hline
$\ell$ & $\tau_{[\mathcal{F}_-^{(\ell-1/2)}]_0}$ & $\lambda_{\s\s Q^+ [\mathcal{F}_-^{(\ell-1/2)}]_0} / \lambda^{\text{GFF}}_{\s\s Q^+ [\mathcal{F}_-^{(\ell-1/2)}]_0}$ &  $\tau_{[\mathcal{B}_+^{(\ell)}]_0}$ & $\lambda_{\s\s [\mathcal{B}_+^{(\ell)}]_0}/\lambda^{\text{GFF}}_{\s\s [\mathcal{B}_+^{(\ell)}]_0} $ \\
\hline
 2 & 0.999612 [1] & 1.11003 [1.12724(8)] & 1.28404 & 0.949494 \\
 4 & 1.07003 & 1.07120 & 1.25687 & 0.944553 \\
 6 & 1.09335 & 1.05826 & 1.24142 & 0.948678 \\
 8 & 1.10590 & 1.05128 & 1.23144 & 0.952376 \\
 10 & 1.11400 & 1.04674 & 1.22440 & 0.955338 \\
 12 & 1.11975 & 1.04347 & 1.21912 & 0.957734 \\
 14 & 1.12411 & 1.04096 & 1.21499 & 0.959716 \\
 16 & 1.12754 & 1.03894 & 1.21166 & 0.961389 \\
 18 & 1.13034 & 1.03727 & 1.20889 & 0.962827 \\
 20 & 1.13267 & 1.03586 & 1.20656 & 0.964081 \\
 22 & 1.13465 & 1.03464 & 1.20456 & 0.965189 \\
 24 & 1.13637 & 1.03356 & 1.20281 & 0.966177 \\
 26 & 1.13786 & 1.03261 & 1.20128 & 0.967066 \\
 28 & 1.13918 & 1.03176 & 1.19992 & 0.967873 \\
 30 & 1.14036 & 1.03099 & 1.19870 & 0.968610 \\
 32 & 1.14142 & 1.03028 & 1.19761 & 0.969286 \\
 34 & 1.14237 & 1.02964 & 1.19661 & 0.969911 \\
 36 & 1.14324 & 1.02905 & 1.19570 & 0.970489 \\
 38 & 1.14404 & 1.02850 & 1.19487 & 0.971028 \\
 40 & 1.14477 & 1.02799 & 1.19410 & 0.971532 \\
 42 & 1.14545 & 1.02751 & 1.19339 & 0.972004 \\
 44 & 1.14608 & 1.02706 & 1.19274 & 0.972448 \\
 46 & 1.14666 & 1.02664 & 1.19212 & 0.972866 \\
 48 & 1.14720 & 1.02625 & 1.19155 & 0.973261 \\
 50 & 1.14771 & 1.02589 & 1.19101 & 0.973657 \\
 \hline
\end{tabular}
\end{center}
 \caption{\label{tab:spectrum} CFT data for the leading $\mathcal{F}_-^{(\ell-1/2)}$ and $\mathcal{B}_+^{(\ell)}$ trajectories in the $\mathcal{N}=1$ super-Ising model computed using the Lorentzian inversion formula. This data corresponds to the $\text{LI}_{z=.05}: \{\mathbb{1}, \s, \e, [\s\s]_{2-98}^{\ell_0=100}, [\s\e]_{0-99}^{\ell_0=100}\}$ computation shown in Figs.~\ref{plot:leadingtwist} and~\ref{plot:leadingOPE}. Rigorous determinations for the stress-tensor data are shown in square brackets.}
\end{table}

Unfortunately, our extremal spectrum OPE coefficient data at $\Lambda = 51$ seemed to fluctuate by $O(1)$ amounts and did not appear reliable. We suspect this is due to a numerical difficulty in resolving the mixing effects along with the sharing effect described in~\cite{Liu:2020tpf}. Further study will be needed in order for us to understand the most reliable way to minimize these effects at high derivative order.

We have also investigated the CFT data using the navigator method~\cite{Reehorst:2021ykw}. In our navigator computations, we imposed the following gaps in the spin-2 sectors: $\Delta_{\mathcal{B}_+^{(2)}} > 3.2$, $\Delta_{\mathcal{B}_-^{(2)}} > 3$, $\Delta_{Q^+ \mathcal{F}_-^{(3/2)}} > 4$, $\Delta_{Q^- \mathcal{F}_+^{(5/2)}} > 4$, in addition to the scalar gaps described in section~\ref{sec:results}; gaps in other channels are the corresponding unitarity bound shifted by $10^{-10}$.  To better understand the data associated to the stress tensor, we computed a rigorous bound on its OPE coefficient across the island at $\Lambda=19$, i.e.~we maximized/minimized $\lambda_{\s \s Q^+ [\mathcal{F}_-^{(3/2)}]_0}$ inside the island.\footnote{We used the $\Sigma$-navigator \cite{Reehorst:2021ykw}, i.e.~the normalization vector is given by summation of a few crossing vectors at discrete $\{\Delta_i,\ell_i\}$ in the channels $\mathcal{B}_+^{(\ell)}$, $\mathcal{F}_-^{(\ell-1/2)}$, $\mathcal{B}_-^{(\ell)}$, $\mathcal{F}_+^{(\ell+1/2)}$. Specifically in each channel where they exist we chose $\ell=0,2,4$ and included 10 $\Delta$s between the gap and the gap+3. However the specific choice of the normalization vector doesn't affect the bound on $\lambda_{\s \s Q^+ [\mathcal{F}_-^{(3/2)}]_0}$.} The result is that the OPE coefficient must live in the range 
\begin{align}
\lambda_{\s \s Q^+ [\mathcal{F}_-^{(3/2)}]_0} / \lambda^{\text{GFF}}_{\s \s Q^+ [\mathcal{F}_-^{(3/2)}]_0} =&\, 1.12724(8)\,, \\
C_T / C_T^{\text{free}} =&\, 1.68414(14)\,.
\end{align}
Comparing with the data in table~\ref{tab:spectrum}, we can see that the analytic calculation of the spin-2 coefficient has an error at the $\sim 1 \%$ level. 

We have also extracted the extremal data from minimizing the navigator function over the island, so far computed up to $\Lambda = 27.$\footnote{We used the same gaps and $\Sigma$-navigator as before. The specific location of the minimum point of the navigator function can vary slightly if we choose a different normalization vector for the $\Sigma$-navigator. However in practice the difference is very small.} As shown in \cite{Reehorst:2021ykw}, the CFT data at the minimum navigator point can give a better estimation, compared to the data at an arbitrary point inside the island. In particular, we extracted the OPE data at the minimum navigator point using the script \texttt{spectrum.py}~\cite{Komargodski:2016auf, spectrum}. The OPE data is included in Fig.~\ref{plot:leadingOPE},\footnote{Here we have dropped a few data points at large spins $>18$ which started randomly fluctuating. In cases where an operator appears at the imposed gap we plot the averaged OPE coefficients in order to reduce the sharing effect. Let us also comment that the scaling dimension data overlaps very closely with the points shown in Fig.~\ref{plot:leadingtwist} so we have not shown them explicitly.} and for the most part it sits between the two most precise analytic curves, giving confidence that the analytic calculations are converging towards physically sensible results.

\begin{figure}[tbp]
  \centering
  \includegraphics[width=.85\textwidth]{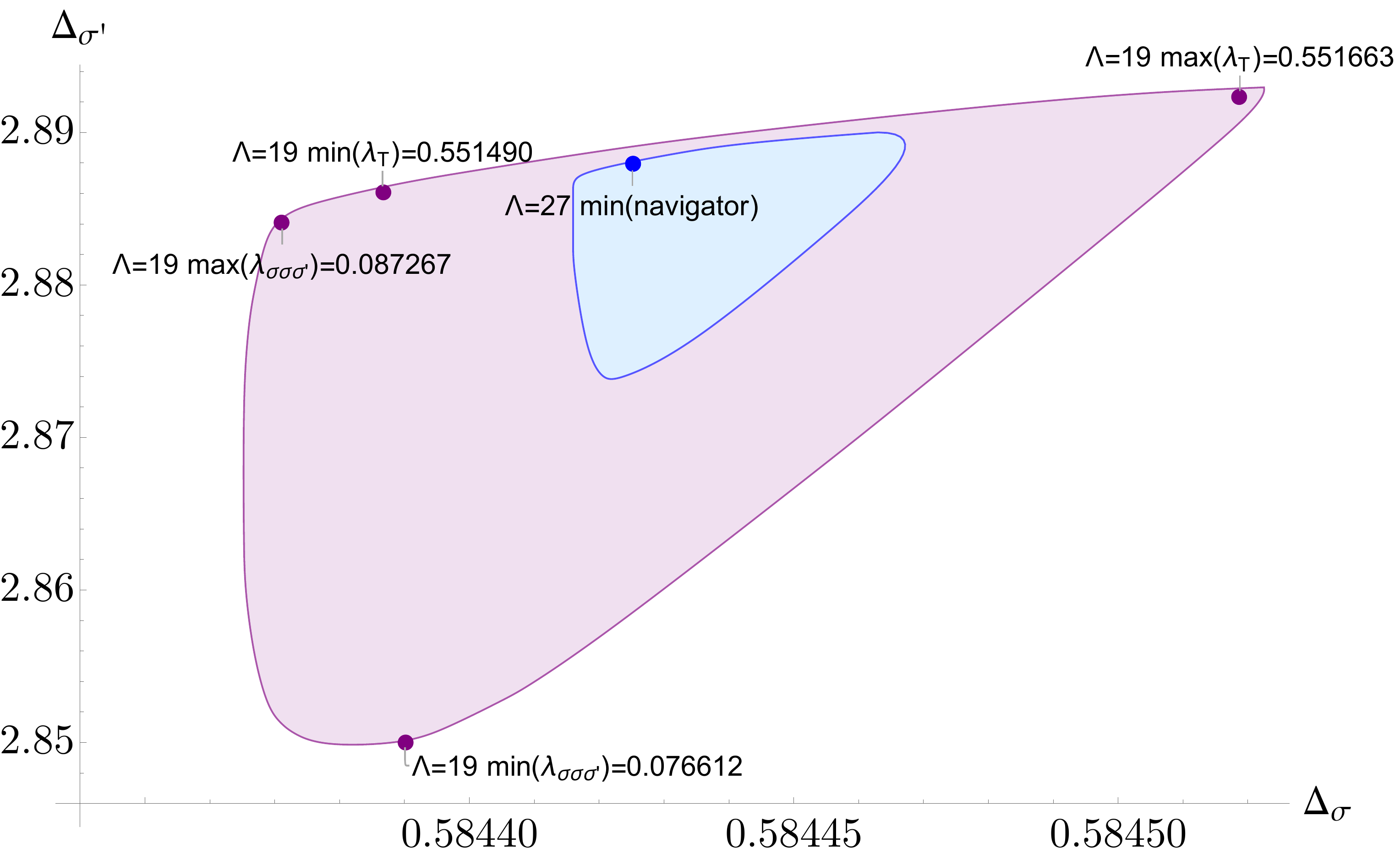}
  \caption{\label{plot:navigatorrun} The locations of the final points from various navigator runs. Purple points: locations where various OPE coefficients are maximized/minimized at $\Lambda=19$. $\lambda_T$ is defined as $\lambda_{\sigma\sigma T}/\Delta_\sigma$. Blue points: the location where navigator value is minimized at $\Lambda=27$. Purple area: $\Lambda=19$ allowed region. Blue area: $\Lambda=27$ allowed region.}
\end{figure}

The specific locations for final points in the navigator computations is shown in Fig.~\ref{plot:navigatorrun}.

In future work it could be helpful to use the numerical bootstrap to compute and incorporate rigorous bounds on the CFT data of the $[\mathcal{B}_+^{(2)}]_0$ and $[\mathcal{B}_-^{(0)}]_1$ multiplets\footnote{A preliminary bound using the navigator at $\Lambda = 19$ is $\lambda_{\s\s Q^2[\mathcal{B}_-^{(0)}]_1} = \lambda_{\s\s\s'} = 0.0819(53)$.} in order to further improve our understanding of the mixing effects at low spin. It will also be important to perform a careful study of the $z$ dependence of the analytic trajectories and explore the extrapolation down to spin 0 and the leading Regge intercepts. We can also incorporate the trajectories with asymptotic twist $\Delta_{\sigma} + 2$ as exchanged operators in order to further improve our results. Nevertheless, it should be clear that the $\mathcal{N}=1$ super-Ising model is an excellent laboratory for further development of analytic bootstrap methods.

\section{Conclusion}
\label{sec:conclusion}
In this work, we have pushed the numerical bootstrap calculation of the critical exponents of the $\mathcal{N}=1$ supersymmetric extension of Ising model up to  $\Lambda = 59$, improving on the $\Lambda=27$ computation performed in~\cite{Rong:2018okz}. 
As can be seen from Figure~\ref{plot:island}, the rate at which the size of the islands shrink as we increase $\Lambda$ has became quite slow at $\Lambda = 59$. To further improve the numerical precision of the critical exponents, na\"ively increasing $\Lambda$ will not be the best strategy, which is quite cost intensive. One obvious thing to attempt is to study mixed correlators containing the superfields $\Sigma$ and $\Sigma'$. Experience from studying mixed correlators of the non-supersymmetric Ising and $O(N)$ vector models tells us that this strategy may lead to considerable improvements in the precision of the numerical calculation. We leave this study for future work. 

We also worked out the analytic bootstrap for the $\mathcal{N}=1$ super-Ising model and obtained precise formulas for the conformal data of super-multiplets that belong to the leading Regge trajectories. In addition to giving a precise picture of the spectrum, our formulas may be useful in future attempts at combining numerical and analytic methods.

It will be interesting to generalize this work to theories with higher supersymmetries. One obvious target is the $\mathcal{N}=2$ super-Ising model studied in \cite{Bobev:2015vsa,Bobev:2015jxa}.  The operator mixing we have encountered here is not unique to $\mathcal{N}=1$ supersymmetric models. This phenomena is in fact quite general in superconformal field theories. For example, superconformal blocks appearing in chiral 4-point functions in $\mathcal{N}=2$ SCFTs take the form
\begin{align}
\mathcal{G}_{\Delta,\ell}(u,v)=G_{\Delta,\ell}(u,v)+a_1 G_{\Delta+1,\ell+1} (u,v)+ a_2 G_{\Delta+1,\ell-1}(u,v)+ a_3 G_{\Delta+2,\ell} (u,v)\,.
\end{align}
The constants $a_i$ are fixed by superconformal symmetry and are given explicitly in \cite{Bobev:2015vsa,Bobev:2015jxa}. 
In case of the $\mathcal{N}=2$ super-Ising model, this superconformal block corresponds to long multiplets that appear in the OPE of a chiral superfield $\Phi$ and an anti-chiral superfield $\Phi^{\dagger}$. 
The four point function of the chiral primaries can be written as 
\begin{align}
\langle \phi(x_1) \phi^{\dagger}(x_2)\phi(x_3)\phi^{\dagger}(x_4) \rangle \sim \sum_{\sc O} f_{\phi\phi^{\dagger} {\sc O}}^2\mathcal{G}_{\Delta_{\sc O},\ell_{\sc O}}(u,v)\,.
\end{align}
Comparing with generalized free fields, an operator with scaling dimension and spin $\{\Delta,\ell\}$ can either be a superconfomal primary of a multiplet with $\{\Delta_{\sc O}, \ell_{\sc O} \}=\{\Delta,\ell\}$ or a superconformal descendant of a multiplet with $\{\Delta_{\sc O}, \ell_{\sc O} \}=\{\Delta-1,\ell-1\}$, and so on. This mixing problem was resolved using supersymmetry relations for $\mathcal{N}=2$ SCFTs in \cite{Gimenez-Grau:2021wiv}. By studying the extremal spectra of the numerical bootstrap kink observed in \cite{Bobev:2015vsa,Bobev:2015jxa}, one might then be able to compare the numerical result and the analytic predictions. Moreover, a SUSY inversion formula was derived in \cite{Gimenez-Grau:2021wiv}. (See also \cite{Alday:2017vkk} for an earlier formula of four dimensional $\mathcal{N}=4$ superconformal field theories.) These SUSY inversion formulas allow the superconformal OPE coefficients $f_{\phi\phi^{\dagger} {\sc O}}^2$, instead of averaged quantities like \eqref{eq:sssscond1}, to be be obtained directly by integrating over the double discontinuity of the crossed channels' correlation functions. It will be interesting to also derive similar SUSY inversion formulas for $\mathcal{N}=1$ theories, making the solution of the operator mixing problem much easier.  We leave such explorations for future work.

\section*{Acknowledgements}
DP was supported by Simons Foundation grant 488651 (Simons Collaboration on the Nonperturbative Bootstrap) and DOE grants DE-SC0020318 and DE-SC0017660. The work of JR was supported by the Deutsche Forschungsgemeinschaft (DFG) through the Emmy Noether research group ``The Conformal Bootstrap Program'' project number 400570283. AA was supported by the Hertz and NDSEG Fellowships. This project has received funding from the European Research Council (ERC) under the European Union’s Horizon 2020 research and innovation programme (grant agreement no. 758903). The computations of this paper were performed on the Yale Grace computing cluster, supported by the facilities and staff of the Yale University Faculty of Sciences High Performance Computing Center, and the Symmetry cluster at the Perimeter Institute.

\appendix
\section{Parameters}
\label{app:parameters}

We used the following choices for the set of spins to compute the islands at each value of $\Lambda$:
\begin{align}\label{eq:spinsets}
S_{35, 43} &= \{0,\dots,44\}\cup \{47, 48, 51, 52, 55, 56, 59, 60, 63, 64, 67, 68\}\,,\nonumber\\
S_{51} &= \{0,\dots,44\}\cup \{47, 48, 51, 52, 55, 56, 59, 60, 63, 64, 67, 68, 71, 72, 75, 76, 79, 80\}\,,\nonumber\\
S_{59} &= \{0,\dots,61\}\cup \{64, 65, 68, 69, 72, 73, 76, 77, 80, 81, 84, 85\}\,. 
\end{align}
The \texttt{sdpb} parameters used in our computations of the bootstrap islands are given in table~\ref{tab:params}, while the parameters used in our computations of the OPE coefficient bounds are given in table~\ref{tab:paramsOPE}.

\begin{table}
\begin{center}
\begin{tabular}{@{}|c|c|c|c|c@{}}
	\toprule
$\Lambda$ &  35, 43 & 51 & 59 \\
{\small\texttt{keptPoleOrder}}& 30 & 30 & 50 \\
{\small\texttt{order}}& 60 & 60 & 120 \\
{\small\texttt{spins}} & $S_{35, 43}$ & $S_{51}$ & $S_{59}$  \\
{\small\texttt{precision}} & 960 & 1024& 1024 \\
{\small\texttt{dualityGapThreshold}} &  $10^{-74}$ & $10^{-74}$  & $10^{-200}$ \\
{\small\texttt{primalErrorThreshold}}&  $10^{-30}$ & $10^{-30}$ & $10^{-200}$ \\
{\small\texttt{dualErrorThreshold}} & $10^{-30}$ & $10^{-30}$ & $10^{-200}$\\ 
{\small\texttt{initialMatrixScalePrimal}} & $10^{60}$& $10^{61}$ & $10^{30}$\\
{\small\texttt{initialMatrixScaleDual}} &  $10^{60}$& $10^{61}$ & $10^{30}$\\
{\small\texttt{feasibleCenteringParameter}} &  0.1 & 0.1 & 0.1\\
{\small\texttt{infeasibleCenteringParameter}} &  0.3 & 0.3 & 0.3\\
{\small\texttt{stepLengthReduction}} &  0.7 & 0.7 & 0.7\\
{\small\texttt{maxComplementarity}} & $10^{201}$ & $10^{201}$ & $10^{200}$\\
 \bottomrule
\end{tabular}
\caption{\label{tab:params}Parameters used for the computations of the conformal bootstrap islands. The sets $S_{\Lambda}$ are defined in (\ref{eq:spinsets}).}
\end{center}
\end{table}

\begin{table}
\begin{center}
\begin{tabular}{@{}|c|c|@{}}
	\toprule
$\Lambda$ & 51 \\
{\small\texttt{keptPoleOrder}}& 32  \\
{\small\texttt{order}}& 60  \\
{\small\texttt{spins}} &  $S'_{51}$   \\
{\small\texttt{precision}} & 1408 \\
{\small\texttt{dualityGapThreshold}} &  $10^{-100}$  \\
{\small\texttt{primalErrorThreshold}}&   $10^{-200}$ \\
{\small\texttt{dualErrorThreshold}} &  $10^{-200}$ \\ 
{\small\texttt{initialMatrixScalePrimal}} & $10^{20}$ \\
{\small\texttt{initialMatrixScaleDual}} &  $10^{20}$ \\
{\small\texttt{feasibleCenteringParameter}} &  0.1 \\
{\small\texttt{infeasibleCenteringParameter}} &  0.3 \\
{\small\texttt{stepLengthReduction}} &  0.7 \\
{\small\texttt{maxComplementarity}} & $10^{200}$\\
 \bottomrule
\end{tabular}
\caption{\label{tab:paramsOPE}Parameters used for the computations of the OPE coefficient bounds. The set $S'_{51}$ is given by $S'_{51} = \{0,\dots,44\}\cup \{47, 48, 51, 52, 55, 56, 59, 60, 63, 64, 67, 68\}$.}
\end{center}
\end{table}

\begin{table}
\begin{center}
\begin{tabular}{@{}|c|c|c|c|c@{}}
	\toprule
$\Lambda$ &  19 & 27 \\
{\small\texttt{keptPoleOrder}}& 14 & 20 \\
{\small\texttt{order}}& 28 & 40  \\
{\small\texttt{spins}} & $S_{19}$ & $S_{27}$  \\
{\small\texttt{precision}} & 500 & 1400 \\
{\small\texttt{dualityGapThreshold}} &  $10^{-20}$ & $10^{-30}$ \\
{\small\texttt{primalErrorThreshold}}&  $10^{-20}$ & $10^{-30}$ \\
{\small\texttt{dualErrorThreshold}} & $10^{-20}$ & $10^{-30}$ \\ 
{\small\texttt{initialMatrixScalePrimal}} & $10^{20}$& $10^{50}$\\
{\small\texttt{initialMatrixScaleDual}} &  $10^{20}$& $10^{50}$\\
{\small\texttt{feasibleCenteringParameter}} &  0.1 & 0.1 \\
{\small\texttt{infeasibleCenteringParameter}} &  0.3 & 0.3\\
{\small\texttt{stepLengthReduction}} &  0.7 & 0.7\\
{\small\texttt{maxComplementarity}} & $10^{100}$ & $10^{800}$\\
 \bottomrule
\end{tabular}
\caption{\label{tab:params_nvg}Parameters used for the navigator computations. The sets $S_{\Lambda}$ are defined in (\ref{eq:spinsets_nvg}).}
\end{center}
\end{table}

For the navigator computations, we used the following choices for the set of spins:
\begin{align}\label{eq:spinsets_nvg}
S_{19} &= \{0,\dots,26\}\cup \{49,50\}\,,\nonumber\\
S_{27} &= \{0,\dots,26\}\cup \{29,30,33,34,37,38,41,42,45,46,49,50\}\,.
\end{align}
The \texttt{sdpb} parameters for the navigator computations are given in table~\ref{tab:params_nvg}. To compute the bound of $\lambda_{\s \s Q^+ [\mathcal{F}_-^{(3/2)}]_0}$ at $\Lambda=19$, we used Algorithm 2 in \cite{Reehorst:2021ykw} with $g_{\text{tol}}=10^{-15}$. To compute the minimum navigator point at $\Lambda=27$, we used Algorithm 1 in \cite{Reehorst:2021ykw} with $g_{\text{tol}}=10^{-15}$. For both navigator computations, to compute the gradient of the navigator function, we used a finite difference of $10^{-20}$ in each argument of the navigator function. The gradient computation can be done using the \texttt{approx\_objective} program in SDPB version 2.5.  

\section{Coefficients}
\label{app:coefficients}
Ratios of OPE coefficients involving different bosonic components of superfields were given in Appendix A of~\cite{Rong:2018okz}. Evaluating them on the dimensions of double-twist operator gives\footnote{The arguments are $[\Delta, \ell]$ for $\mathcal{B}^{(\ell)}_{\pm}$ multiplets and $[\Delta + 1/2, \ell - 1/2]$ for $\mathcal{F}^{(\ell)}_{\pm}$ multiplets.}
\begin{align}
\frac{\lambda_{\e\e[\mathcal{B}^{(\ell)}_+]_{n}}}{\lambda_{\s\s[\mathcal{B}^{(\ell)}_+]_{n}}} &= c_1[2\ds + 2n + \ell,\ell] = \frac{n(2n+2\ell +1)}{\ds(2\ds-1)}\,,\\
\frac{\lambda_{\e\e Q^2 [\mathcal{B}^{(\ell)}_-]_{n}}}{\lambda_{\s\s Q^2[\mathcal{B}^{(\ell)}_-]_{n}}} &= d_1[2\ds + 2n + \ell - 1, \ell] = \frac{(2\ds + n -2)(4\ds + 2n + 2\ell - 3)}{\ds(2\ds -1 )}\,,\\
\frac{\lambda_{\e\e Q^{+} [\mathcal{F}^{(\ell - 1/2)}_-]_{n}}}{\lambda_{\s\s Q^{+} [\mathcal{F}^{(\ell - 1/2)}_-]_{n}}} &= e_1[2\ds + 2n + \ell, \ell - 1] = -\frac{n(4\ds + 2n + 2\ell - 3)}{\ds (2\ds - 1)}\,,\\
\frac{\lambda_{\e\e Q^{-} [\mathcal{F}^{(\ell + 1/2)}_+]_{n}}}{\lambda_{\s\s Q^{-} [\mathcal{F}^{(\ell + 1/2)}_+]_{n}}} &= f_1[2\ds + 2n + \ell, \ell] = -\frac{(2n + 2 \ell+1) (2 \ds +n-2)}{\ds (2 \ds -1)} \,,\\
\left(\frac{\lambda_{\s\e Q^2[\mathcal{B}^{(\ell)}_+]_{n}}}{\lambda_{\s\s [\mathcal{B}^{(\ell)}_+]_{n}}}\right)^2 &= c_2[2\ds + 2n + \ell, \ell] = \frac{(2 \ds +2 n-1) (\ds + n + \ell) (2 \ds +2n + \ell-1)}{2 \ds  (2 \ds -1) (4 \ds +4n + 2 \ell-1)}\,,\\
\left(\frac{\lambda_{\s\e[\mathcal{B}^{(\ell)}_-]_{n+1}}}{\lambda_{\s\s Q^2 [\mathcal{B}^{(\ell)}_-]_{n+1}}}\right)^2 &= d_2[2\ds + 2n + \ell + 1, \ell] = \frac{2 (\ds +n) (2 \ds +2n + 2 \ell+1) (4 \ds +4n+2 \ell+1)}{\ds  (2 \ds -1) (2 \ds + 2n + \ell)}\,,\\
\left(\frac{\lambda_{\s\e Q^-[\mathcal{F}^{(\ell - 1/2)}_-]_{n}}}{\lambda_{\s\s Q^+[\mathcal{F}^{(\ell - 1/2)}_-]_{n}}}\right)^2 &= e_2[2\ds + 2n + \ell, \ell-1] = \frac{\ell (2 \ds +2 n-1) (2 \ds + 2n + 2 \ell-1)}{2 \ds  (2 \ds -1) (2 \ell-1)}\,,\\
\left(\frac{\lambda_{\s\e Q^+[\mathcal{F}^{(\ell - 3/2)}_+]_{n+1}}}{\lambda_{\s\s Q^-[\mathcal{F}^{(\ell - 3/2)}_+]_{n+1}}}\right)^2 &= f_2[2\ds + 2n + \ell, \ell - 2] = \frac{\ell (2 \ds +2 n-1) (2 \ds + 2n + 2 \ell-1)}{2 \ds  (2 \ds -1) (2 \ell-1)}\,.
\end{align}
To include anomalous dimensions one can replace $n \rightarrow n+\delta h_{\mathcal{O}}$.

\clearpage
\bibliography{Biblio}
\bibliographystyle{utphys}
\end{document}